\documentclass[11pt]{article}
\usepackage{amssymb}
\usepackage{geometry}                
\usepackage{graphicx}
\usepackage{rawfonts}
\usepackage{amssymb}
\usepackage{epstopdf}
\usepackage{amsmath,amsthm,amssymb}
\usepackage{mathtools}
\usepackage{fullpage}
\usepackage{color}
\usepackage[usenames,dvipsnames,svgnames]{xcolor}
\usepackage[colorlinks=true,citecolor=blue,linkcolor=red,urlcolor=blue]{hyperref}
\usepackage[font=footnotesize, labelfont=bf, margin=0.5cm]{caption}
\usepackage[labelformat=simple]{subcaption}
\usepackage{rawfonts}
\usepackage{etex}
\usepackage{placeins}

\newtheorem{theo}{Theorem}

\newtheorem{cor}{Corollary}

\def\Pr{\noindent \emph{Proof: }}
\def\qed{$\Box$}

\usepackage{graphicx,color}
\usepackage{amsmath,amssymb,latexsym}
\usepackage{mathrsfs}

\def\sfrac#1#2{\hbox{\scalebox{0.8}{$\displaystyle \frac{#1}{#2}$}}}
\def\Ref#1{(\ref{#1})}

\input{prepictex}
\input{pictex}
\input{postpictex}

\usepackage{ulem}

\begin{document}

\title{Thermodynamic and topological properties of copolymer rings with a segregation/mixing transition}

\author{E J Janse van Rensburg$^1$, E Orlandini$^2$, M C Tesi$^3$ and S G Whittington$^4$ \\
$^1$  Department of Mathematics and Statistics, York University,\\ Toronto, Ontario M3J~1P3, Canada \\
$^2$ Dipartimento di Fisica e Astronomia e Sezione INFN, Universit\`a di Padova, \\ Via Marzolo 8, I-35131 Padova, Italy \\
$^3$ Dipartimento di Matematica, Universit\`a di Bologna, \\ Piazza di Porta San Donato 5, I-40126 Bologna, Italy \\
$^4$ Department of Chemistry, University of Toronto, Toronto M5S 3H6, Canada \\}
\date{\today}
\vspace{10pt}

\date{\today}
\vspace{10pt}

\maketitle

\begin{abstract} 
Two ring polymers close to each other in space may be either in a segregated phase
if there is a strong repulsion between monomers in the polymers, or intermingle in a 
mixed phase if there is a strong attractive force between the monomers.  These phases 
are separated by a critical point which has a $\theta$-point character.  The metric and 
topological properties of the ring polymers depend on the phase, and may change abruptly
at the critical point.  In this paper we examine the thermodynamics and linking
of two ring polymers close in space in both the segregated and mixed phases using
a cubic lattice model of two polygons interacting with each other.  Our
results show that the probability of linking is low in the segregated phase, but that
it increases through the critical point as the model is taken into the mixed phase.
We also examine the metric and thermodynamic properties of the model, with focus
on how the averaged measures of topological complexity are related to these properties.
\end{abstract}


\section{Introduction}
\setcounter{equation}{0}

Mutually attracting pairs of circular or ring polymers in solution undergo a transition from 
a segregated to a mixed phase at a critical temperature.  The metric and topological 
properties of the polymers are different in the segregated and mixed phases, changing 
at the critical temperature from expanded spatially segregated conformations to 
more compact conformations in the mixed phase where the two polymers interpenetrate. 
In~\cite{DeGennes:1979} the segregation-mixing transition of a polymer-polymer-solvent
mixture is discussed in chapter IV.4.   There it is argued that in a good solvent the polymer 
coils behave like hard spheres which cannot interpenetrate (they are segregated).  
In a poor solvent, however, the coils tend to exclude the solvent and are driven 
together increasing the local concentration which, if high enough, should drive
the system through a $\theta$-transition into a collapsed phase.  This should 
also occur if there is a strong attractive interaction between the polymer coils,
where they exclude solvent molecules by mixing in close proximity to one another.

\begin{figure}[t!]
  \centering
  \includegraphics[width=0.85\textwidth]{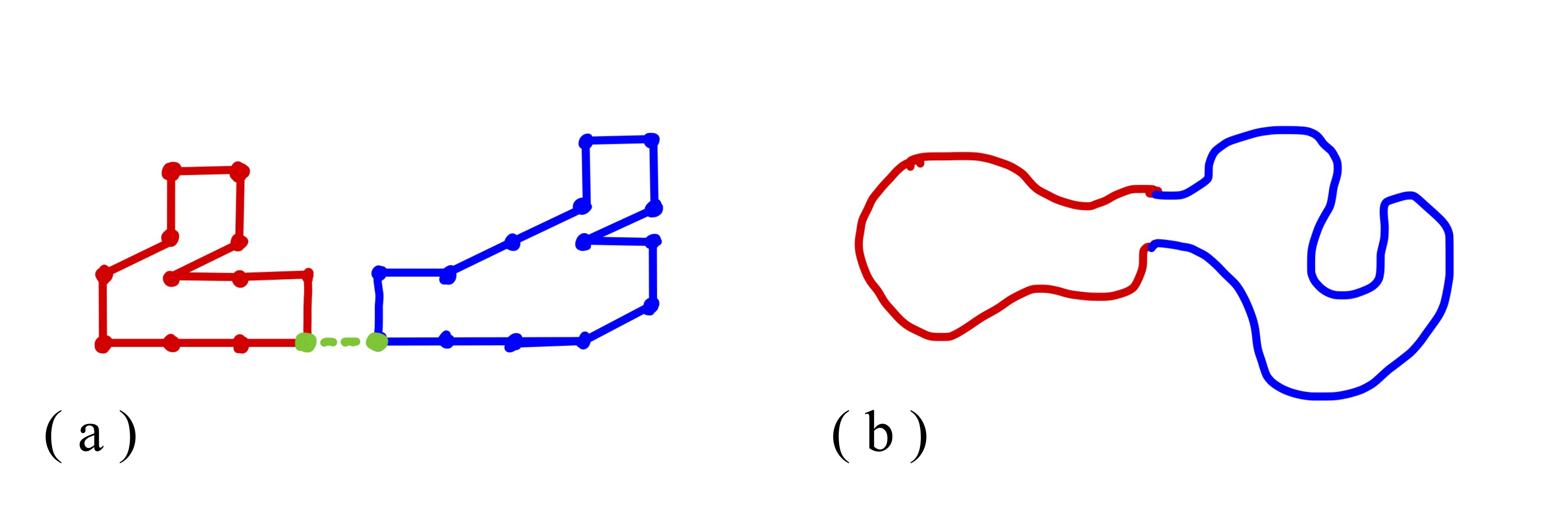}
  \caption{(a) The model. The dashed line joining a vertex in each component is rooted
  (that is, it  cannot be moved by an elementary move in the Monte Carlo algorithm. 
 (b) A diblock copolymer ring with the ends of the two blocks forced to be nearest 
 neighbours  in space.  This approximates a figure eight conformation with each
 polygon ring a block in the copolymer.}
\label{fig:0}
\end{figure}

In this paper we aim to model the segregated-mixed phases in a model of a system 
composed of a pair of proximate ring polymers which may be linked, especially when 
they are in the mixed  phase.  We use a cubic lattice closed self-avoiding walk model 
(lattice polygons) where the proximity is modelled by forcing the polygons to have at 
least one pair of vertices (one vertex in each polygon) a unit distance apart.  
See figure~\ref{fig:0}(a) for an example.  The two polygons are both self-avoiding and 
mutually avoiding in the lattice.

Our model is also useful as a model for a particular class of diblock copolymers with figure 
eight connectivity composed of two polygons joined together by sharing a single step
(see figure~\ref{fig:0}b).  Each polygon is a block in the copolymer, and both polygons are 
in a good solvent but there is a short range interaction \textit{between} vertices located
in each of the two polygons.  If the interaction is a strong attractive force, then the two 
polygons in the figure eight will tend to interpenetrate, otherwise they will segregate due to an 
entropic repulsion between them.  One may also consider ring formation in a uniform
$4$-star polymer with two $A$-arms and two $B$-arms. The star polymer can be cyclized 
to form an $A$ ring and a $B$ ring, in conditions with different interaction strengths between 
the two rings, resulting in different extents of linking of the two rings.

We shall focus (in particular) on the metric and topological properties of our model in
the segregated and mixed phases.  Our aim is to address the following questions:
\begin{itemize}
\item How do configurational properties (such as mean extension, shape and entanglement) 
depend on the strength of the interaction between the component polygons (or blocks)?
\item How do topological properties of the system (such as the complexity of linking between
the component polygons) differ in the segregated and mixed phases?
\item How does topological complexity, as measured in terms of the link spectrum, 
depend on the degree of mixing?
\end{itemize}
The configurational and topological properties of the model could be different in the 
segregated and mixed phases.  In the segregated phase the component polygons tend 
to be separated in space, and so the degree of linking between them will be low, while 
in the mixed phase they may be strongly intermingled with a high degree of linking.  
It is not clear \textit{a priori} that the transition will be strongly signalled in the 
configurational properties of the model.  While the segregated phase will have the scaling
properties of a self-avoiding walk, as the model crosses the critical point into the
mixed phase the two polygons might form ribbon-like
structures~\cite{van1994lattice,janse1996entanglement}.
If this is all that happens then 
the critical exponents of the mixed phase will still be given by self-avoiding walk 
exponents.  The interactions between different parts of the ribbon boundaries could lead to further
collapse, similar to that of a self-avoiding walk having gone through a 
$\theta$-transition into its dense phase.  This transition should be signalled in the 
linking (and in the mean topological invariants such as linking number) of the two components, and
the critical properties (such as the radius of gyration exponent) might be different on the two sides of the transition.

The plan of the paper is as follows.  In section \ref{sec:model} we describe the model
and its partition function and free energy. 
Section \ref{sec:rigorous} contains some theorems establishing the existence of a transition 
from a segregated phase to a mixed phase.  In section \ref{sec:results_thermo} we discuss
the Monte Carlo methods used to sample configurations and consider the results 
of the simulations (and in particular the metric and shape properties of the system 
as a function of the attractive interaction).
Results on the topological mutual entanglement are presented and discussed in 
section \ref{sec:entangle}.  We close with a short Discussion in section \ref{sec:discussion}.

\section{The model}
\label{sec:model}

We model circular polymers as lattice self-avoiding polygons of length $n$.  These are 
embeddings of simple closed curves in the simple cubic lattice $\mathbb{Z}^3$.  The 
embeddings are also simple closed curves in $\mathbb{R}^3$ with well defined topological 
properties (knots and links).  In this paper the lattice polygons are mutually avoiding 
and placed in $\mathbb{Z}^3$ such that a pair of vertices (one vertex in each polygon) 
are a unit distance apart (see figure~\ref{fig:0}).  We label the two polygons by 
$A$ and $B$ respectively so that this is a model of two adjacent rings $A$ and $B$ which 
may be linked, or parts of a copolymer with a figure eight connectivity and two blocks, 
each a ring of the figure eight.

A \textit{mutual contact} between polygons $A$ and $B$ is a pair of vertices $(v_A,v_B)$, 
such that $v_A\in A$ and $v_B\in B$ and the distance between $v_A$ and $v_B$ is equal 
to one:  $d(v_A,v_B)=1$.  An interaction between the two polygons (or blocks in the 
copolymer) is introduced by introducing an energy $\epsilon_m$ associated with each mutual 
contact, and then defining the parameter $\beta_m = -\epsilon_m/k_BT$ where
$k_B$ is Boltzmann's constant and $T$ is the absolute temperature.  Denoting 
the number of conformations of the lattice model of total length $2n$ (each polygon 
component of length $n$) with $k_m$ mutual contacts by $p^{(2)}_{2n}(k_m)$,  
the equilibrium properties  of the system are given by the partition function
\begin{equation}
Z_{2n}(\beta_m)= \sum_{k_m} p^{(2)}_{2n}(k_m) e^{\beta_m k_m} .
\label{eq_Z}
\end{equation}
Entropy dominates the model when $\beta_m \le 0 $ and the two component polygons 
tend to stay separated (this is a \textit{segregated phase} due to the mutual avoidance 
between the polygons inducing a short-ranged repulsion between them).  When 
$\beta_m>0$ is large enough, then the mutual contacts induce an attraction between 
the component polygons, and one expects this to increase the number of mutual contacts.  
In this case one expects the most likely conformations to be those where the two polygons 
interpenetrate strongly in a \textit{mixed phase}.

The free energy per unit length is given by $f_{2n}(\beta_m)
=\sfrac{1}{2n} \log Z_{2n}(\beta_m)$ and by taking $n\to\infty$ the limiting free energy 
of the model is obtained:
\begin{equation}
f(\beta_m) = \lim_{n\to\infty}\frac{1}{2n} \log Z_{2n}(\beta_m) .
\label{eq_f}
\end{equation}
It is not known that this limit exists for all values of $\beta_m\in\mathbb{R}$ but we shall 
show that it exists for $-\infty < \beta_m \le 0$ and is equal to $\kappa_3=\log \mu_3$
where $\kappa_3$ is the connective constant and $\mu_3$ is the growth constant 
of cubic lattice self-avoiding walks~\cite{hammersley1961number} 
(see section~\ref{sec:rigorous}).  When $\beta_m >0$ the situation is more complicated 
and we shall rely on Monte Carlo simulations to explore the model in this regime.

\section{Some rigorous results}
\label{sec:rigorous} 

In this section we obtain some bounds on the partition function $Z_{2n}(\beta_m)$ and 
use these to prove the existence of the thermodynamic limit when $\beta_m \le 0$ and 
the existence of a phase transition in the system.  We also obtain some similar results 
for fixed link type.

Attach the coordinate system $(x_1,x_2,x_3)$ to $\mathbb{Z}^3$ so that the coordinates 
of the vertices are all integers.  Write $p_n$ for the number of $n$-edge polygons (modulo 
translation) so that $p_{2m+1}=0$, $p_4=3$, $p_6=22$, etc.  
Hammersley~\cite{hammersley1961number} has shown that 
$\lim_{n\to\infty} \sfrac{1}{n} \log p_n = \log \mu_3$ where $\mu_3$ is the growth 
constant of self-avoiding walks on this lattice.  Similarly, if we write $q_n$ for the 
corresponding number of polygons on the square lattice $\mathbb{Z}^2$ then 
$\lim_{n\to\infty} \sfrac{1}{n} \log q_n = \log \mu_2$ where $\mu_2$ is the growth 
constant of self-avoiding walks on the square lattice.

\begin{theo}
If $-\infty < \beta_m \le 0$ then 
$\displaystyle \lim_{n\to\infty} \sfrac{1}{2n} \log Z_{2n} (\beta_m)
= \kappa_3 = \log \mu_3$.
\label{thm1}
\end{theo}
\Pr
To obtain an upper bound consider $\beta_m=0$.  Embed a polygon with $n$ edges 
in $p_n$ ways and embed a second $n$-edge polygon in a box of side $2n$ centred 
on the first polygon,  This implies that $Z_{2n}(0) \le p_n^2 e^{o(n)}$ so that 
\[ \limsup_{n\to\infty} \frac{1}{2n} \log Z_{2n}(\beta_m) 
\le \limsup_{n\to\infty} \frac{1}{2n} \log Z_{2n}(0)    \le \log \mu_3,
\quad \hbox{if $\beta_m \le 0$.} \]
To get a lower bound  construct two polygons, one ($\sigma_1$) with $n{-}2$ edges 
and the other ($\sigma_2$) with $n$ edges.  For $\sigma_1$ translate the right-most
top-most edge a unit distance to the right and add two edges to reconnect the polygon 
to form $\sigma_3$.  Translate $\sigma_3$, and rotate it if necessary, so that its right-most 
edge is unit distance from an edge of $\sigma_2$ and there are exactly two vertices 
of $\sigma_3$ that are unit distance from vertices of $\sigma_2$.  This gives the bound
\[ Z_{2n} (\beta_m) \ge p_n p_{n-2}\, e^{2\beta_m}/2 \]
since exactly two new mutual contacts are created.
Taking logarithms, dividing by $2n$ and letting $n\to\infty$ gives
\[  \liminf_{n\to\infty}  \frac{1}{2n}  \log Z_{2n} (\beta_m) \ge \log \mu_3 \]
which completes the proof.
\qed

\begin{theo}
If $\beta_m > 0$ and $0 < \alpha <1$ then 
\[ \liminf_{n\to\infty}  \sfrac{1}{2n} \log Z_{2n} (\beta_m) 
\ge \sfrac{\alpha}{2}\left[ \log \mu_2 + \beta_m \right] + (1-\alpha) \log \mu_3. \]
Since $\alpha\in (0,1)$ is arbitrary, this shows that
\[ \liminf_{n\to\infty}  \sfrac{1}{2n} \log Z_{2n} (\beta_m) 
\ge \max\left( \sfrac{1}{2}\left[ \log \mu_2 + \beta_m \right],\log \mu_3  \right) . \]
\label{thm2}
\end{theo}
\Pr 
Construct a polygon $\sigma_1$ in the plane $x_1=0$ with $\lfloor \alpha n\rfloor$ edges, 
$0<\alpha < 1$.  Let $\sigma_2$ be a translate of $\sigma_1$ in the plane 
$x_1=1$.  Let $\sigma_3$ be a polygon in ${\mathbb Z}^3$ with $n-\lfloor\alpha n\rfloor$
edges and with no vertices with $x_1 > -1$. Similarly, let $\sigma_4$  be a polygon 
in ${\mathbb Z}^3$ with $n-\lfloor\alpha n\rfloor$ edges and with no vertices with $x_1< 2$.  
Concatenate $\sigma_1$ and $\sigma_3$ to obtain a polygon with $n$ edges 
and, similarly, concatenate $\sigma_2$ and $\sigma_4$.  The two resulting 
polygons have $\lfloor \alpha n \rfloor$ pairs of vertices unit distance apart so that they 
contribute a Boltzmann factor  $\exp[\beta_m \lfloor\alpha n\rfloor]$ to the partition 
function $Z_{2n}(\beta_m)$.  $\sigma_1$ can be chosen in 
$\mu_2^{\lfloor\alpha n\rfloor+o(n)}$ ways, $\sigma_2$ can be chosen in only one way, 
while $\sigma_3$ and $\sigma_4$ can each be chosen in 
$\mu_3^{n - \lfloor\alpha n\rfloor + o(n)}$ ways.  This construction gives a 
lower bound on $Z_{2n}(\beta_m)$ \[ Z_{2n}(\beta_m) \ge q_{\lfloor\alpha n\rfloor}\, 
p_{n-\lfloor\alpha n\rfloor}^2 e^{\beta_m \lfloor\alpha n\rfloor}/4 
= \mu_2^{\lfloor\alpha n\rfloor + o(n)} \mu_3^{2(n-\lfloor\alpha n\rfloor) + o(n)} 
e^{\beta_m \lfloor \alpha n \rfloor} .\]
Taking logarithms of the above, dividing by $2n$, and letting $n \to \infty$, the 
claimed lower bound is obtained. 
\qed

\begin{theo}
$\displaystyle \liminf_{n\to\infty}  \sfrac{1}{2n} \log Z_{2n} (\beta_m)$ is 
non-analytic at $\beta_m^0$ where 
$\displaystyle 0 \le \beta_m^0 \le 2\log \mu_3 - \log \mu_2$.
\label{thm3}
\end{theo}

\Pr
We can rewrite the result of theorem 2 as 
\[ \liminf_{n\to\infty} \sfrac{1}{2n}\log Z_{2n} (\beta_m) 
\ge \alpha\left[ \sfrac{1}{2} \left( \log \mu_2 + \beta_m \right)
- \log \mu_3\right] + \log \mu_3 \]
and this is equal to $\log \mu_3$ when 
\[\sfrac{1}{2} \left( \log \mu_2 + \beta_m \right) - \log \mu_3 = 0, 
\quad  \hbox{for $0<\alpha <1$.} \]
This shows that the limiting infimum is greater than $\log \mu_3$ when $\beta_m > 
2 \log \mu_3 - \log \mu_2$.  Therefore the limiting infimum is
singular at some $\beta_m^0$ where 
\[ 0 \le \beta_m^0 \le 2\log \mu_3 - \log \mu_2 \]
which completes the proof. \qed

\subsection{Bounds on the free energies of linked conformations}

The free energies of linked conformations of specified links can also be bounded using 
arguments similar to the above.  We proceed by recalling that the growth constant of 
unknotted polygons is defined by the limit \cite{Pippenger:1989:DAM}
\begin{equation}
\lim_{n\to\infty} \sfrac{1}{n} \log p_n(\emptyset) = \log \mu_\emptyset.
\end{equation}
It is known that 
$\mu_\emptyset < \mu_3$~\cite{Pippenger:1989:DAM,Sumners&Whittington:1988:J-Phys-A}.  
One may similarly define the growth constant $\mu_K$ of knotted polygons of knot type $K$ by
\begin{equation}
\limsup_{n\to\infty} \sfrac{1}{n} \log p_n(K) = \log \mu_K .
\end{equation}
It is known that $\mu_\emptyset \leq \mu_K < \mu_3$~\cite{soteros1992entanglement}

\begin{figure}[h!]
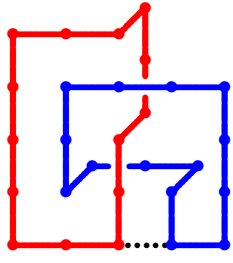

\beginpicture
\setcoordinatesystem units <2pt,2pt>
\setplotarea x from -70 to 60, y from -10 to 60

\setplotsymbol ({\scalebox{0.50}{$\bullet$}})

\color{black}
\setdots <3.5pt>
\plot 20 0 30 0 /

\setsolid
\setplotsymbol ({\scalebox{0.50}{$\bullet$}})

\color{red}
\plot 0 0 20 0 20 20 25 25 25 28 /
\plot 25 32 25 45 20 40 /
\plot 20 40 0 40 0 0 /
\multiput {$\bullet$} at 0 0 10 0 20 0 20 10 20 20 25 25 25 35 25 45
20 40 10 40 0 40 0 30 0 20 0 10  /

\color{blue}
\plot 10 10 15 15 18 15 /
\plot 22 15 35 15 30 10 30 0 40 0 40 30 10 30 10 10 /
\multiput {$\bullet$} at 10 10 15 15 25 15 35 15 30 10 30 0 40 0 
40 10 40 20 40 30 30 30 20 30 10 30 10 20  /

\color{black}
\normalcolor
\endpicture
\caption{A linked conformation of two polygons with one mutual contact shown.  
The total number of mutual contacts between the two polygons is $k_m=13$, 
while each polygon component has length $n=14$.  In this case the link type is 
$L=2_1^2$, the Hopf link.}
\label{fig:A}
\end{figure}

Denote the number of linked conformations of link type $L$, with polygon components
of length $n$ each, and with $k_m$ mutual contacts, by $p_{2n}^{(2)}(k_m,L)$.
The partition function is 
\begin{equation}
Z_{2n}^{(2)} (\beta_m,L) = \sum_{k_m} p_{2n}^{(2)}(k_m,L)\,e^{\beta_m k_m}
\end{equation}
and it is a sum of weighted conformations of fixed linked type $L$ and total length $2n$.

An example of a linked conformation in our model is shown in figure \ref{fig:A}.
This conformation is a Hopf-link and it has weight $e^{13 \beta_m}$ (since
$k_m=13$) in the partition function.  More generally, the two components of a lattice link 
of link type $L$ are lattice knots of knot types $K_1$ and $K_2$ respectively.   In many cases 
$K_1=K_2=\emptyset$ (for example, if $L$ is the Hopf link, as shown in figure \ref{fig:A}), 
but $K_1$ and $K_2$ could be (necessarily) non-trivial knots for certain link types, or could 
be chosen to be given knot types.

We generalize theorem \ref{thm1} as follows.

\begin{theo}
Suppose that $L$ is a 2-component link with components of knot types $K_1$ and $K_2$.
Then, if $-\infty < \beta_m \leq 0$, 
\[ \log \mu_\emptyset 
\leq \liminf_{n\to\infty} \sfrac{1}{2n} \log  Z_{2n}^{(2)} (\beta_m,L)
\leq \limsup_{n\to\infty} \sfrac{1}{2n} \log Z_{2n}^{(2)} (\beta_m,L)
< \log \mu_3 .\]
In the event that $K_1=K_2=\emptyset$, then 
$\displaystyle \lim_{n\to\infty} \sfrac{1}{2n} \log Z_{2n}^{(2)} (\beta_m,L)
= \log \mu_\emptyset < \log \mu_3$.
\label{thm4}
\end{theo}

\Pr
An upper bound is obtained by first noting that $Z_{2n}^{(2)} (\beta_m,L)
\leq Z_{2n}^{(2)} (0,L)$, and then counting the number of conformations
of the component polygons independently, times the number of ways they 
may be placed so that the link may be recovered.  Each component polygon
is a placement of a simple closed polygon of fixed knot type, say $K_1$
for the first polygon, and $K_2$ for the second polygon. This shows that
\[ Z_{2n}^{(2)} (1,L) \leq n^3\,p_n(K_1)\,p_n(K_2) \]
Taking logarithms, dividing by $2n$ and letting $n\to\infty$, gives
\[ \limsup_{n\to\infty} \sfrac{1}{2n} \log Z_{2n}^{(2)} (\beta_m,L)
\leq \sfrac{1}{2}(  \log \mu_{K_1} + \log \mu_{K_2} )
< \log \mu_3 . \]
In the event that the two component polygons of the link are both
the unknot, then $K_1 = K_2 = \emptyset$ and
\[ \limsup_{n\to\infty} \sfrac{1}{2n} \log Z_{2n}^{(2)} (\beta_m,L)
\leq \log \mu_\emptyset . \]

\begin{figure}[h!]
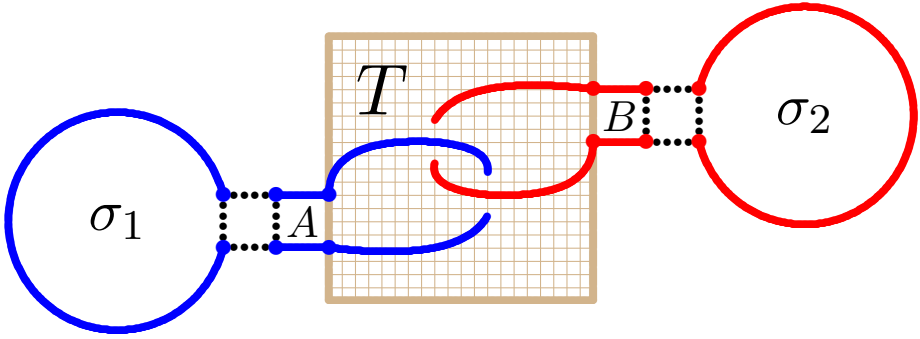

\beginpicture
\setcoordinatesystem units <2pt,2pt>
\setplotarea x from -30 to 200, y from -10 to 60

\setcoordinatesystem units <1pt,1pt>
\setplotarea x from 100 to 200, y from 0 to 100
\setplotsymbol ({\scalebox{2}{.}})
\color{Tan} 
\grid 20 20

\setcoordinatesystem units <2pt,2pt>
\setplotarea x from -10 to 200, y from -10 to 60
\setplotsymbol ({\scalebox{0.66}{$\bullet$}})
\color{black}
\setdots <4pt>
\plot 30 10 40 10 40 20 30 20 30 10 /
\plot 110 30 120 30 120 40 110 40 110 30 /

\setsolid
\color{Tan}
\plot 50 0 100 0 100 50 50 50 50 0  /
\color{black}
\setquadratic
\color{blue}
\plot 50 20 55 28 70 30 78 28 80 24 /
\plot 50 10 70 10 80 16 /
\setlinear
\plot 40 10 50 10 / \plot 40 20 50 20 /
\multiput {\scalebox{1.33}{$\bullet$}} at 50 10 40 10 50 20 40 20 /
\circulararc 330 degrees from 30 20 center at 10 15
\multiput {\scalebox{1.33}{$\bullet$}} at 30 10 30 20 / 

\setquadratic
\color{red}
\plot 100 40 80 40 70 34 /
\plot 70 26 72 22 80 20 95 22 100 30 /

\setlinear
\plot 110 30 100 30 / \plot 110 40 100 40 /
\multiput {\scalebox{1.33}{$\bullet$}} at 110 30 100 30 110 40 100 40 /
\circulararc 330 degrees from 120 30 center at 140 35
\multiput {\scalebox{1.33}{$\bullet$}} at 120 30 120 40 / 

\color{black}
\put {\scalebox{2}{$\sigma_1$}} at 10 15
\put {\scalebox{2}{$\sigma_2$}} at 140 35
\put {\scalebox{2.5}{$T$}} at 60 40
\put {\scalebox{1.5}{$A$}} at 45 15
\put {\scalebox{1.5}{$B$}} at 105 35

\color{black}
\normalcolor
\endpicture
\caption{A schematic diagram with a  construction to create a link in the cubic lattice.
The link is represented by a tangle $T$ drawn inside a square in the 
$x_3=0$ plane of the lattice.  By subdividing the cubic lattice, the tangle 
can be pushed onto the edges in the $x_3=0$ plane  and only stepping 
into the $x_3=1$ plane to create overpasses in the projection.  Moreover,
one component of the tangle has endpoints in the left-most boundary of
the square while the other has endpoints in its right-most boundary. 
By adding edges at $A$ and $B$
respectively, the components of $T$ can be closed into polygons such 
that the bottom and top edges of $T$ are in different components.
The top edge of an unknotted polygon $\sigma_1$ is concatenated 
to the bottom edge of $T$ at $A$, and the bottom edge of an unknotted
polygons $\sigma_2$ is concatenated to the top edge of $T$ at $B$.}
\label{fig:B}
\end{figure}

Consider the schematic diagram in figure \ref{fig:B} to find the lower bound.  
A link $L$ is represented as a two-dimensional tangle $T$ which is embedded 
in the $x_3=0$ plane of the cubic lattice, accommodating overpasses in $T$ by 
stepping into the $x_3=1$ plane.  The embedding of $T$ is completed into a link by
closing each component into a polygon such that the edge with lexicographic
least midpoint (the \textit{bottom edge}), and the edge with lexicographic
most midpoint (the \textit{top edge}), are in different components of
the tangle.  This is shown as $A$ and $B$ in figure \ref{fig:B}.

Proceed by concatenating the top edge of an unknotted polygon $\sigma_1$ onto 
the bottom edge $A$, and the bottom edge of a second unknotted polygon $\sigma_2$ 
onto the top edge $B$, as illustrated. Assume that the length of the first component 
in the embedded tangle $T$ is $m_1$, and of the second component, $m_2$.  
Fix the length of $\sigma_1$ to be $n{-}m_1$, and of $\sigma_2$ to be $n{-}m_2$.  
This creates a link $L$ of link type determined by $T$, and since the overpasses 
in $T$ are accommodated by overstepping into the $x_3=1$ plane, there is
at least one mutual contact between the components of $T$, as required.  
There are $j_m\geq 1$ mutual contacts in $L$, and $j_m\leq 4(m_1{+}m_2)$ since 
a polygon of length $m_1$ has at most $4$ mutual contacts for each vertex, and 
since $\sigma_1$ and $\sigma_2$ do not contribute any such mutual contacts.

Since the orientation of the top edge of $\sigma_1$ has to match
that of $A$, there are $p_{n-m_1}(\emptyset)/2$ choices for $\sigma_1$.
Similarly, there are $p_{n-m_2}(\emptyset)/2$ choices for $\sigma_2$.
This shows that
\[ p_{n-m_1}(\emptyset)\,p_{n-m_2}(\emptyset)\, e^{\beta_m\,j_m} 
\leq 2^2\, Z_{2n}^{(2)} (\beta_m,L) . \]
Take logarithms, divide by $2n$, and let $n\to\infty$.  Since $m_1$ and $m_2$
are fixed, this shows that
\[ \log \mu_\emptyset \leq  \liminf_{n\to\infty} \sfrac{1}{2n} \log
 Z_{2n}^{(2)} (\beta_m,L) . \]
 This completes the proof. \qed

\begin{figure}[h]
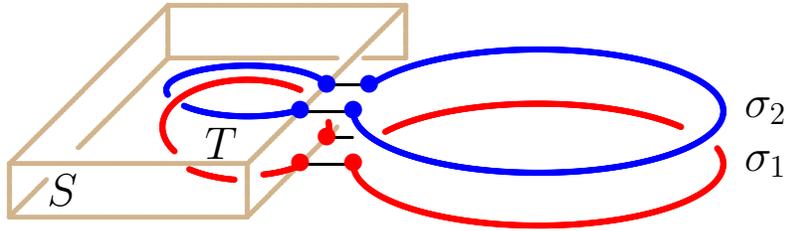

\beginpicture
\setcoordinatesystem units <2pt,2pt>
\setplotarea x from -100 to 100, y from -15 to 35

\setplotsymbol ({\scalebox{1.5}{.}})
\color{Tan}
\plot -15 -10 2 7 /
\plot -15 0  15 30 /
\plot -15 0 -60 0 -30 30 15 30 /
\plot -15 -10 -60 -10 / \plot -60 -10 -53 -3 /  \plot -47 3 -30 20 2 20 / \plot 7 20 15 20 /
\plot -15 -10 -15 0 /
\plot -60 -10 -60 0 /
\plot -30 20 -30 30 /
\plot 15 20 15 30 /

\setplotsymbol ({\scalebox{2}{.}})

\color{Blue}
\ellipticalarc axes ratio 3:1 337 degrees from 5 10 center at 40 10
\color{Red}
\ellipticalarc axes ratio 3:1 195 degrees from 5 0 center at 40 0
\ellipticalarc axes ratio 3:1 110 degrees from 67 7 center at 40 0

\color{Blue}
\ellipticalarc axes ratio 3:0.9 180 degrees from 0 15 center at -15 14
\ellipticalarc axes ratio 3:1 93 degrees from -27 11 center at -15 14 

\color{Red}
\ellipticalarc axes ratio 1.8:1 20 degrees from 0 5 center at -15 7 
\ellipticalarc axes ratio 1.8:1 160 degrees from -5 14 center at -15 7 
\ellipticalarc axes ratio 1.8:1 30 degrees from -26 -1 center at -15 7 
\ellipticalarc axes ratio 1.8:1 30 degrees from -12 -2 center at -15 7 

\setplotsymbol ({\scalebox{1}{.}})
\color{Black}
\plot -5 0 5 0 /  \plot 0 5 5 5 /
\plot -5 10 5 10 /  \plot 0 15 8 15 /

\color{Blue}
\multiput {\LARGE$\bullet$} at 0 15 -5 10 8 15 5 10 /

\color{Red}
\multiput {\LARGE$\bullet$} at 0 5  -5 0  5 0 /

\color{Black}
\normalcolor

\put {\LARGE$S$} at -50 -5
\put {\LARGE$T$} at -20 4
\put {\LARGE$\sigma_2$} at 83 10
\put {\LARGE$\sigma_1$} at 83 0
\endpicture
\caption{Schematic of a tangle $T$ embedded in a slab $S$ and  
 two polygons $\sigma_1$ and $\sigma_2$, each of length $n$, 
concatenated onto the components of the tangle.  Since $\sigma_2$ is a translation of
$\sigma_1$ one step along the $x_3$-direction, the number of mutual contacts
between them is $n$.}
\label{fig:6}
\end{figure}

Lower bounds on the free energies of linked conformations of the model in 
figure~\ref{fig:0}(a) are determined using a construction similar to that in the proof 
of theorem \ref{thm2}.  A schematic diagram is shown in figure \ref{fig:6}.  
The intersection of the $x_3=0$ and $x_3=1$ planes and
${\mathbb Z}^3$ is a \textit{slab} $S$ of height $1$.  (That is, $S$ consists of two
square lattice planes a distance one apart in the $x_3$-direction).  Let $T$ be
a tangle diagram of a link of type $L$.  Then $T$ can be realised as two self-avoiding
walks in $S$ such that the endpoints of the self-avoiding walks are in a plane
$x_1=k$, and with the self-avoiding walks confined to the lattice points
in $S$ with $x_1\leq k$.  We next translate the tangle, and extend the endpoints 
of its component self-avoiding walks by adding steps, if necessary, into 
the $x_1>k$ sublattice such that the two component self-avoiding walks have
the same lengths $\ell$, and the the four endpoints have coordinates $(m,0,0)$
$(m,0,1)$, $(m,1,0)$ and $(m,1,1)$ where $(m,0,0)$ and $(m,1,0)$ are
the endpoints of one self-avoiding walk, and $(m,0,1)$ and $(m,1,1)$ are
the endpoints of the second self-avoiding walk.  By adding two edges to
close off the tangle into a linked pair of polygons, the link $L$ is realised
as a lattice link of type $L$. We assume that there are $c_0$ mutual
contacts between the component self-avoiding walks. 

As shown in figure \ref{fig:6}, let $\sigma_1$ be a square lattice
self-avoiding polygon in the $x_3=0$ plane, and $\sigma_2$ be the
translate of $\sigma_1$ one step in the $x_3$-direction. If the length of 
$\sigma_1$ is $n$, then there are $n$ mutual contacts between the
pair of polygons $(\sigma_1,\sigma_2)$.  This pair can be translated
together and concatenated on the link $L$ by placing the left-most and nearest
edge (the bottom edge) of $\sigma_1$ one step in the $x_1$-direction 
to the edge joining the endpoints $(m,0,0)$ and $(m,1,0)$.  Then $\sigma_2$ 
has its bottom edge one step in the $x_1$-direction from the edge 
joining the endpoints $(m,0,1)$ and $(m,1,1)$.  The concatenation
gives a lattice link of type $L$, with total length $2\,\ell{+}2\,n{+}2$.
The total number of mutual contacts is $c_0{+}n$.  This shows that
\begin{equation}
Z_{2\,\ell{+}2\,n{+}2}^{(2)}(\beta_m,L) \geq q_n \, \beta_m^{c_0+n},
\label{eqn6}
\end{equation}
since the number of choices for $\sigma_1$ is $q_n$, the number of
square lattice polygons of length $n$.

By taking logarithms of equation \Ref{eqn6}, diving by $2n$ and then taking
$n\to\infty$, the following theorem is proven.

\begin{theo}
If $\beta_m > 0$ then
$\displaystyle \liminf_{n\to\infty} \sfrac{1}{2n} \log Z_{2n}^{(2)}(\beta_m,L)
\geq \sfrac{1}{2} \left( \log \mu_2 + \beta_m \right)$.
\hfill\qed
\label{thm5}
\end{theo}

The corollary of theorems \ref{thm2} and \ref{thm5} is that there is, for
some link types, a critical point $\beta_m^{(L)}$.

\begin{cor}
Suppose that $L$ is a link with components each of knot type the unknot.
Then the limit $\liminf_{n\to\infty} \sfrac{1}{2n} \log Z_{2n}^{(2)}(\beta_m,L)$ is
non-analytic at a critical point $\beta_m^{(L)}$, where
$\displaystyle 0 \leq \beta_m^{(L)} \leq 2 \log \mu_\emptyset{-}\log\mu_2$.
\label{cor1}
\end{cor}

\Pr
Note that if $\beta_m\leq 0$, then by theorem \ref{thm4},
$\lim_{n\to\infty} \sfrac{1}{2n} \log Z_{2n}^{(2)}(\beta_m,L) = \log \mu_\emptyset$.
On the other hand, if $\beta_m>0$, then by theorem \ref{thm5},
$\liminf_{n\to\infty} \sfrac{1}{2n} \log Z_{2n}^{(2)}(\beta_m,L) \geq 
\sfrac{1}{2} \left( \log \mu_2 + \beta_m \right)$.  This lower bound is
greater than $\log \mu_\emptyset$ if $\beta_m>2\log\mu_\emptyset{-}\log \mu_2$.
Thus, there is a non-analyticity at a critical point 
$\beta_m^{(L)} \in [0,2\log\mu_\emptyset{-}\log \mu_2]$
in $\liminf_{n\to\infty} \sfrac{1}{2n} \log Z_{2n}^{(2)}(\beta_m,L)$
as claimed. \qed

Since $\mu_\emptyset < \mu_3$ the upper bound in corollary \ref{cor1}
is strictly smaller than the upper bound given in theorem \ref{thm3}.
In addition, note that the probability of seeing a link of type $L$
(and with both components the unknot), is 
\begin{equation}
P_{2n}(L) = \frac{Z_{2n}^{(2)}(\beta_m,L)}{Z_{2n}(\beta_m)} .
\end{equation}
By theorems \ref{thm1} and \ref{thm4}, 
\begin{equation}
P_{2n}(L) = K\, \left(\frac{\mu_\emptyset}{\mu_3}(1+o(1)) \right)^{2n}
\longrightarrow 0,\quad\hbox{if $n\to\infty$ and $\beta_m\leq 0$.}
\end{equation}
Since 
$|\mu_3 - \mu_\emptyset| \approx 10^{-6}$~\cite{van2002probability,baiesi2010entropic}, 
the convergence of $P_{2n}(L)$ to zero is numerically very slow and not significant 
until walks have lengths of $O(10^6)$, this effect will not be visible in our data, and 
the probability of linking as a particular fixed link type $L$ will be a function of the 
local geometry of the polygons in the lattice and the value of $\beta_m$.

\section{Results: Thermodynamic and metric properties}
\label{sec:results_thermo}

\subsection{Monte Carlo method}

Conformations of the lattice model were sampled from the 
Boltzmann distribution using a Markov chain Monte Carlo algorithm.  The elementary moves
were a combination of pivot moves for self-avoiding polygons~\cite{madras1990monte}
and local Verdier-Stockmayer style moves~\cite{verdier1962monte}.  The Verdier-Stockmayer 
moves were introduced to increase the mobility of the Markov Chain when the algorithm 
samples at large positive values of $\beta_m$~\cite{tesi1996monte,tesi1996interacting} 
where there is a strong interaction between the component polygons which reduces 
the success rates of the pivot moves.

Sampling was also improved by implementing the elementary moves using a 
Multiple Markov Chain algorithm with chains distributed along a sequence of 
parameters $(\beta_m^{(j)})$ for $j=1,2,\ldots,M$.  Along each parallel chain 
Metropolis sampling was implemented to sample from the Boltzmann distribution 
at a fixed $\beta_m^{(j)}$, and chains were swapped using the protocols of
Multiple Markov Chain sampling~\cite{Geyer:1991:CSS,tesi1996monte,tesi1996interacting}.  
The collection of parallel multiple Markov chains is itself a Markov Chain with stationary 
distribution the product of the Boltzmann distributions along each chain 
(see reference~\cite{tesi1996monte,tesi1996interacting}). 

\begin{figure}[t!]
  \centering
  \includegraphics[width=0.75\textwidth]{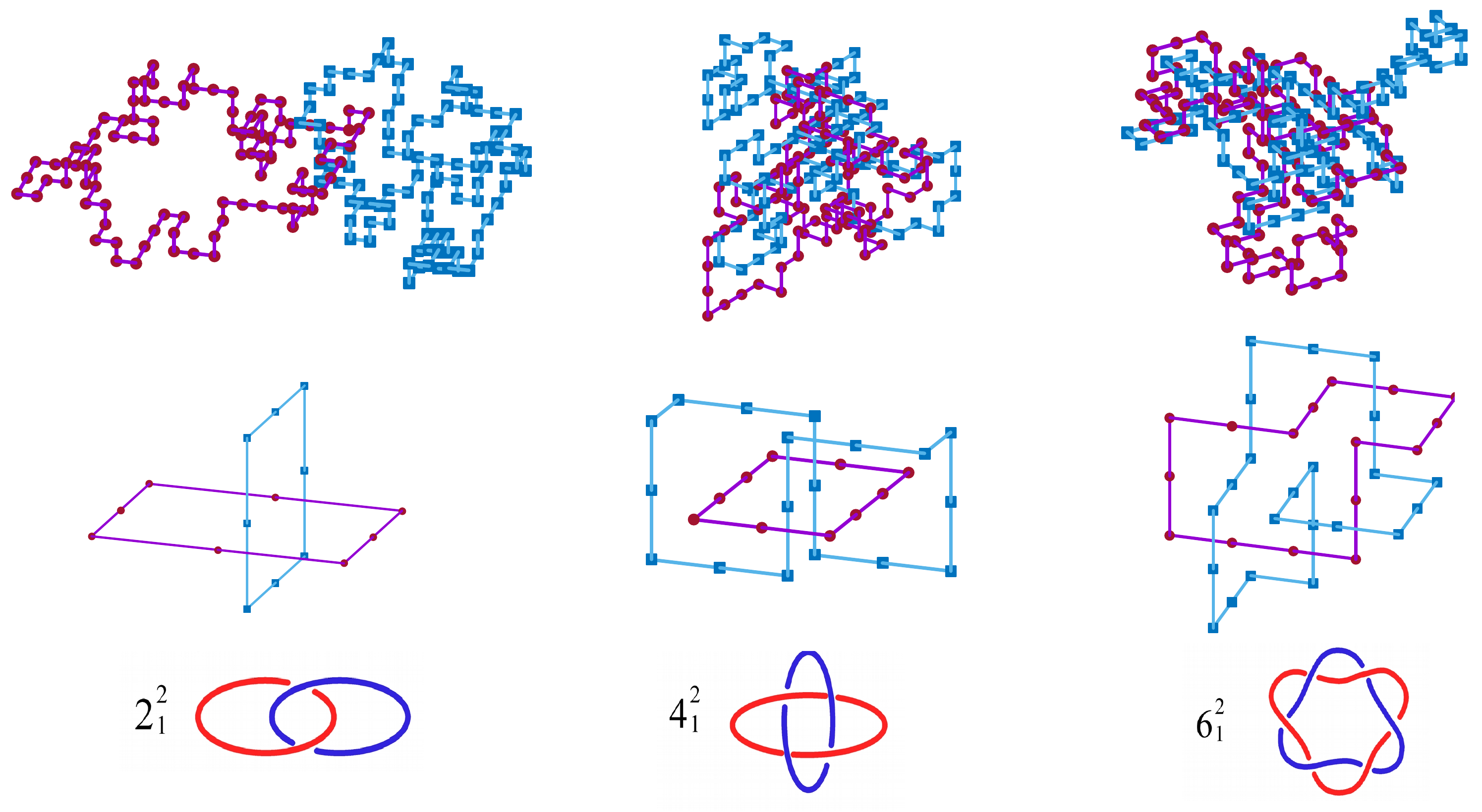}
  \caption{ Top row: Pairs of lattice polygons, each of $n=100$ steps sampled 
  at different values of $\beta_m$'s (0.0, 0.4,0.6). Their corresponding, 
  topologically equivalent, simplified versions after the smoothing and shrinking 
  procedure via the BFACF algorithm are reported in the middle row while the 
  minimal representation of the associated link type is reported in the bottom row.}
\label{fig:1}
\end{figure}

In this paper we sampled along $M \approx 50$ parallel chains, and we were able to obtain
sufficiently uncorrelated samples for systems of \textit{total} size $2n\in \{96,200,296,400,600,800\}$
and for the $\beta_m^{(j)}$ equally spaced in the interval $[-0.3,1.0]$.  We counted
a single iteration as $O(1)$ pivot moves, and $O(n)$ local Verdier-Stockmayer style
moves.  By spacing the reading of data along each Markov chain, we were able to
sample approximately $2.5\times 10^4$ conformations that are essentially 
\textit{uncorrelated} at each fixed value of $\beta_m^{(j)}$ for a total of at least 
$1.25 \times 10^6$ data points for each value of $n$.  The simulations were expensive 
in terms of CPU time.  For example, for $2n=600$ and $M\approx 50$ a total of $25,\!000$ 
uncorrelated conformations were sampled over three months of CPU time.  In 
figure~\ref{fig:1} we show some examples of conformations taken along the chains 
for different values of $\beta_m^{(j)}$.

\begin{figure}[t!]
  \centering
  \includegraphics[width=0.75\textwidth]{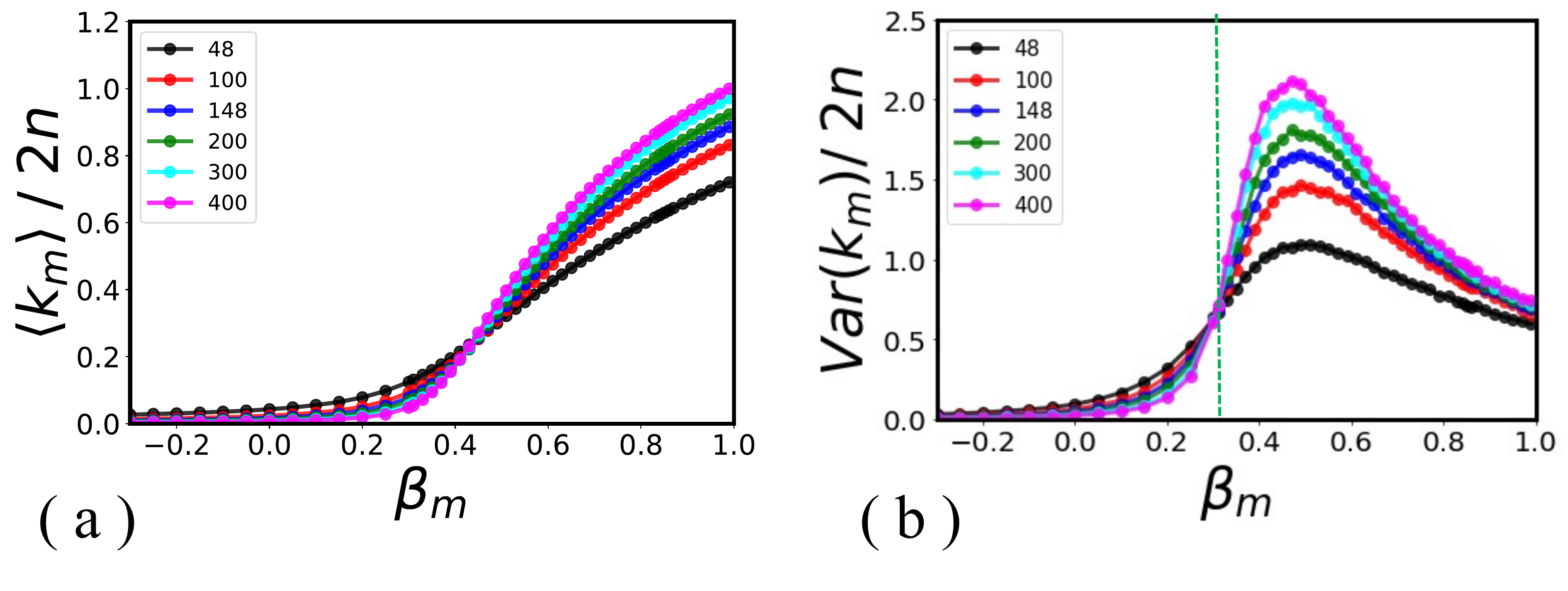}
  \caption{
(a) Plot of the average number of mutual contacts per monomer, 
$\langle k_m\rangle/2n$ and (b) the corresponding variance $Var(k_m)/\; 2n$ as 
a function of $\beta_m$.  Different symbols refer to different values of $n$ (see legend). 
In panel (b) the dashed vertical line highlights the location of the most asymptotic
crossings that we consider as the best estimate of the transition}
\label{fig:2}
\end{figure}

\subsection{Thermodynamic properties}

The results in section \ref{sec:model} show that there should be two regimes
in the model, namely a segregated regime for negative or small  positive values of
$\beta_m$, and a mixed regime of interpenetrating components when
$\beta_m$ is large and positive.  The segregated regime is characterized by
states where the two polygons are separated from one another with a low
density of mutual contacts between them, while the mixed phase has the 
two components close together in the same local space so that the number of
mutual contacts is increased. 

A sharp change in the average energy per monomer $\langle k_m\rangle/(2n)$ is consistent with 
the two regimes being separated by a phase boundary (see section \ref{sec:rigorous}).  This is also seen in the 
variance of $\langle k_m\rangle$ defined by
\begin{equation}
Var(k_m) =  \langle k_m^2\rangle - \langle k_m \rangle^2.
\end{equation} 

Estimates of $\langle k_m\rangle/(2n)$ and the normalized variance $Var(k_m)/(2n)$ are plotted as functions of
$\beta_m$ for various values of $n$ in figure~\ref{fig:2}(a) and in
figure~\ref{fig:2}(b), respectively.
For $\beta_m < \beta_m^*$, $\langle k_m \rangle/(2n)$,  tends to zero with increasing $n$. In this phase
the small and decreasing number of mutual contacts per monomer is consistent with
the two component polygons being largely segregated in space.
This is the \textit{segregated phase}, as explained
in section \ref{sec:rigorous}.  For $\beta_m>\beta_m^*$ the curves of
$\langle k_m\rangle/(2n)$ are increasing as $n$ increases.
In this \textit{mixed phase} the two polygons have large non-zero energy per
monomer (that is, a high incidence of mutual contacts), consistent with the two
component polygons having sections near to each other in the lattice as they
share the same volume in space. Our data are consistent with
$\langle k_m \rangle \to A\,n$ as $n$ increases with $2<A\leq 3$.  This shows that
the mixed phase is not an extended ribbon with the two polygons forming its
boundary, but is a denser phase where strands in each polygon have a high
number of contacts per monomer (between $2$ and $3$) with the
other component.
The segregated-mixed transition as $\beta_m$ is taken through its critical value
is also seen in the peak forming in  $Var(k_m)/2n$  when plotted as a function of $\beta_m$, see  figure~\ref{fig:2}(b).
With increasing $n$
the peaks move to smaller values of $\beta_m^{(p)}$.  This behaviour is consistent
with theorems \ref{thm1} and \ref{thm2}, since in the infinite $n$ limit the variance
is equal to zero in the segregated phase, has a jump discontinuity at the
critical point, and then decreases with increasing $\beta_m$.

The data in figure~\ref{fig:2}(b) strongly suggest that the transition at $\beta_m^*$ is
asymmetric (that is, in the limit $n\to\infty$ the variance is zero if $\beta<\beta_m^*$
but it characteristically increases as $\beta\searrow \beta_m^*$). In these
circumstances the intersections of the variances for different values of $n$
in figure~\ref{fig:2}(b) are a good estimator of the location of the critical point
as $n\to\infty$.  This gives the estimate of $\beta_m^* = 0.31 \pm 0.01$.  The
asymmetry of the transition is consistent with the results of section \ref{sec:rigorous}.

\subsection{Metric and shape properties}
\label{sec:mutual_entangl}

\begin{figure}[t!]
  \centering
  \includegraphics[width=0.75\textwidth]{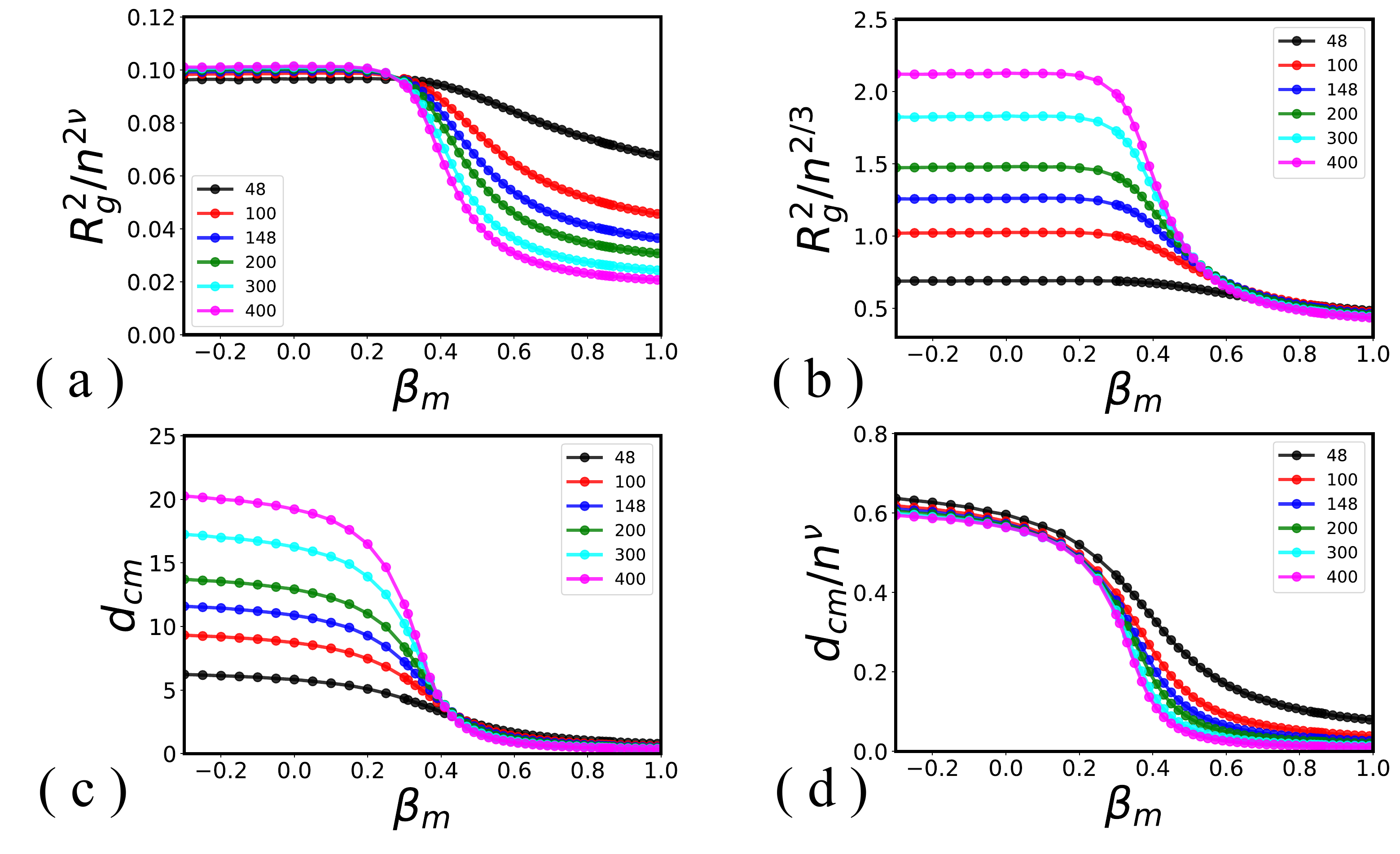}
  \caption{(a,b) $\langle R_g^2\rangle/n^{2\nu}$ for $\nu=\nu_{SAW}$~(a) 
and $\nu=1/3$~(b).  (c) Average distance between the centres of mass of 
the two rings $\langle d_{cm}\rangle$ as a function of $\beta_m$. 
(d) Same as (c) but now scaled by $n^{\nu}$ with $\nu=\nu_{SAW}$. 
Different symbols refer to different values of $n$ (see legend).}
\label{fig:3}
\end{figure}

The size and metric scaling of lattice links can be examined by calculating 
metric quantities such as the mean square radius of gyration $R_j^2$ for
$j\in\{1,2\}$ for components $j=1$ or $j=2$.  As expected  the behaviour of 
$R_j^2$ does not depend on $j$ and we improve the estimate  of this metric 
observable by averaging over the two components:  $R_g^2 = (R_1^2+R_2^2)/2$. 
This quantity, scaled by the power $n^{2\nu}$  is reported, as a function of 
$\beta_m$, in figure~\ref{fig:3}(a) and (b), and for lengths 
$n\in\{48,100,148,200,300,400\}$.  Since the metric scaling of the self-avoiding
walk has exponent $\nu_{SAW}=0.587297(7)$~\cite{Clisby:2010:Phys-Rev-Lett} 
it should be the case that the ratio $R_g^2/n^{2\nu_{SAW}}$ is a constant for $\beta_m<0$ 
(in the segregated regime). This is seen in figure~\ref{fig:3}(a) where the data for 
$\beta \leq \beta_m^* \approx 0.3$ collapse to a constant close to $0.1$, with little 
dependence on $n$. This is evidence that for these values of $\beta_m$ the system is 
in a \textit{segregated phase} where the self-avoidance between
the two polygons separates them in space, and each polygon has the properties
of a ring polymer in a good solvent, with associated metric exponent $\nu$.

For values of $\beta_m > \beta_m^*$ the model is instead in a mixed phase.
Here the ratio $R_g^2/n^{2\nu}$ with $\nu=\nu_{SAW}$ is dependent on both $n$ and $\beta_m$,
decreasing either with increasing $\beta_m$ or with increasing $n$, as seen in 
figure~\ref{fig:3}(a).  The collapsed nature of the model is exposed by plotting 
$R_g^2 /n^{2/3}$ against $\beta_m$, showing collapse of the data for 
different values of $n$ to an underlying curve for large values of $\beta_m$, see
figure~\ref{fig:3}(b). These observations are consistent with the model passing 
through a phase transition into a mixed and \textit{collapsed} phase where 
the interpenetrating components explore states with a high (local) density of 
monomers.  Note that in this figure the critical point $\beta_m^*$ separating 
the segregated and mixed phases has value approximately $0.3$, that is, 
consistent with the one estimated using the variance of the mutual number of 
contact (see figure~\ref{fig:2}(b)).

The collapse of the data for large $\beta_m>\beta_m^*$
is consistent with the two polygons interpenetrating each other in a phase
with high mutual contacts.  This indicates that the mixed phase 
may be characterized by compact conformations in a collapsed phase 
and that the lattice link transitions through a $\theta$-point at $\beta_m^*$ 
from an expanded and segregated phase into a collapsed and mixed phase. 

The transition  between a segregated (and expanded or free) phase for negative 
$\beta_m$, and a mixed (and collapsed) phase for large positive $\beta_m$, is also 
suggested by other metric observables.  For instance, in figure~\ref{fig:3}(c) and (d) 
the mean separations between the centres of mass $d_{cm}$ of the 
two polygon components are examined.  The $\beta_m$ dependence of this measure 
is reported in figure~\ref{fig:3}(c)  while in figure~\ref{fig:3}(d) the scaled version
$d_{cm}/n^\nu$ is plotted.  In both cases the data decrease with increasing 
$\beta_m$, consistent with the model entering a compact phase where the 
centres of mass of the two components are close to each other.  Note that in 
(c) $d_{cm}$ increases with  $n$ in the segregated phase, and this growth is 
shown in (d) to be at the expected rate of $O(n^\nu)$, the typical length scale of 
the model.  The curves in (c) intersect pairwise close to a critical value 
$\beta_m^* \approx 0.4$, slightly larger than the estimate suggested by 
the data of figure~\ref{fig:3}(a), but not inconsistent with the expected 
segregated-mixed transition. 

\begin{figure}[t!]
  \centering
  \includegraphics[width=0.85\textwidth]{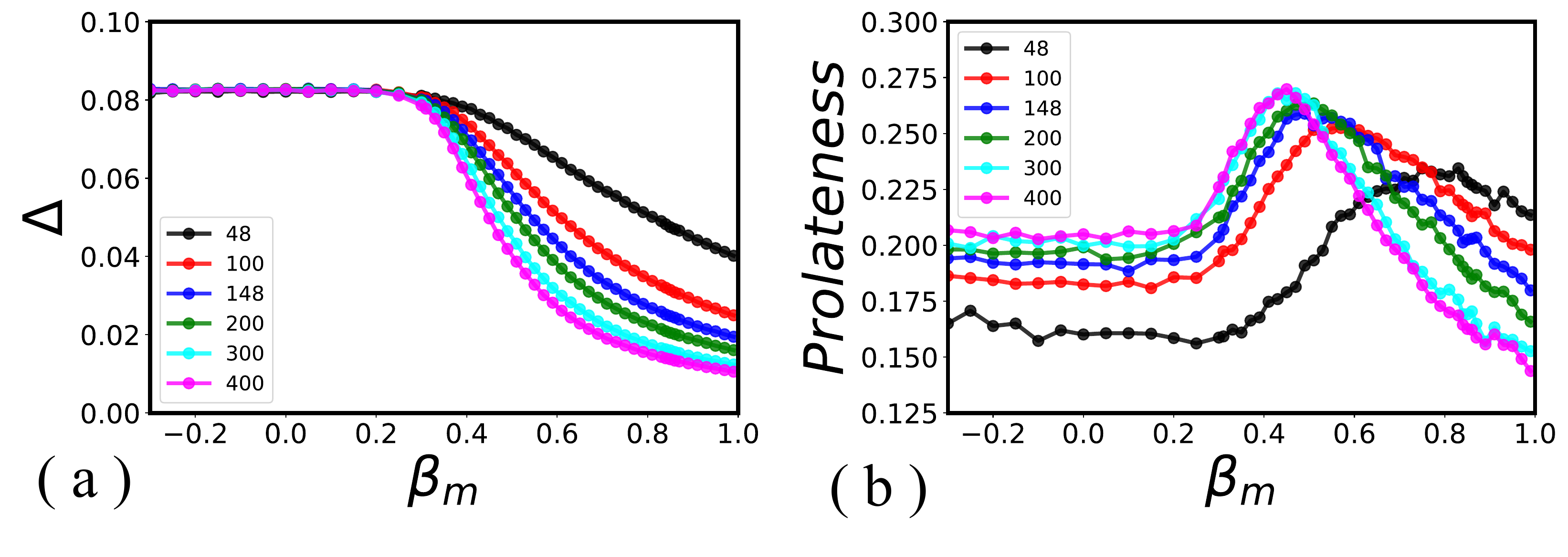}
  \caption{
Asphericity $\Delta$ (a) and degree of prolateness (b) as a function of $\beta_m$. 
Note that the value of $\beta_m$ at which $\Delta$ deviates from the constant value 
characterizing the segregated phase and the most asymptotic position of the peaks 
in the prolateness (i.e. $n=400$) are consistent with the value of $\beta_m^*$.}
\label{fig:4}
\end{figure}

The configurational properties of the model change as it crosses over from
the segregated phase to the mixed phase.  The interaction between the two 
polygon components, due to both the self-avoidance repulsion, and the short
ranged interaction induced by weighted mutual contacts, deform the components
in the two phases, and this may be seen by measuring the asphericity and prolateness
of components. In the segregated phase the conformations may be similar to
that shown in figure~\ref{fig:1} (left), while the mixed phase has interpenetrated
components as shown in figure~\ref{fig:1} (middle and right).  In the segregated
phase the polygon components are aspherical when the components are segregated,
but transitioning into the mixed phase reduces the degree of asphericity
as the two components collapse by forming mutual contacts and interpenetrate
into a locally dense conformation.  This is seen in figure~\ref{fig:4}(a) where the 
average asphericity $\Delta = (\Delta_1+\Delta_2)/2$ is highest in the 
segregated phase but decreases once the model 
transitions through a critical value of $\beta_m$ into the mixed phase.  In the 
segregated phase the data are collapsed into a horizontal line (independent of both 
$\beta_m$ and $n$) with the asphericity $\Delta\approx 0.08$ over the entire range 
of the segregated phase).  In the mixed phase $\Delta$ decreases with increasing 
$\beta_m$ and increasing $n$.

The degree of prolateness of the model (plotted in figure~\ref{fig:4}(b)) presents 
a more nuanced picture, and our data are more noisy.  They suggest a relatively 
constant (in $\beta_m$)  value in the segregated phase that increases and appears 
to peak in the mixed phase at a location that moves to smaller values of $\beta_m$ 
with increasing $n$ approaching the expected value $\beta_m^*$.   When 
$\beta_m \le 0$ the main effect is entropic repulsion where the two curves are mutually 
repelling (for entropic reasons), leading to the two components being prolate.  
As $ \beta_m$ increases towards its critical value the two components will start to 
intermingle but there will still be parts of each component that are not intermingled 
(so that the components are not completely mixed).  These intermingled parts might 
feel a stronger entropic repulsion resulting in a more prolate shape.  For large $\beta_m$ 
the mutual attraction dominates, and this will overcome the entropic repulsion and give 
a more spherical shape when the two components are mixed.  Clearly, from the figure, 
this effect is small.

\begin{figure}[ht!]
  \centering
  \includegraphics[width=0.425\textwidth]{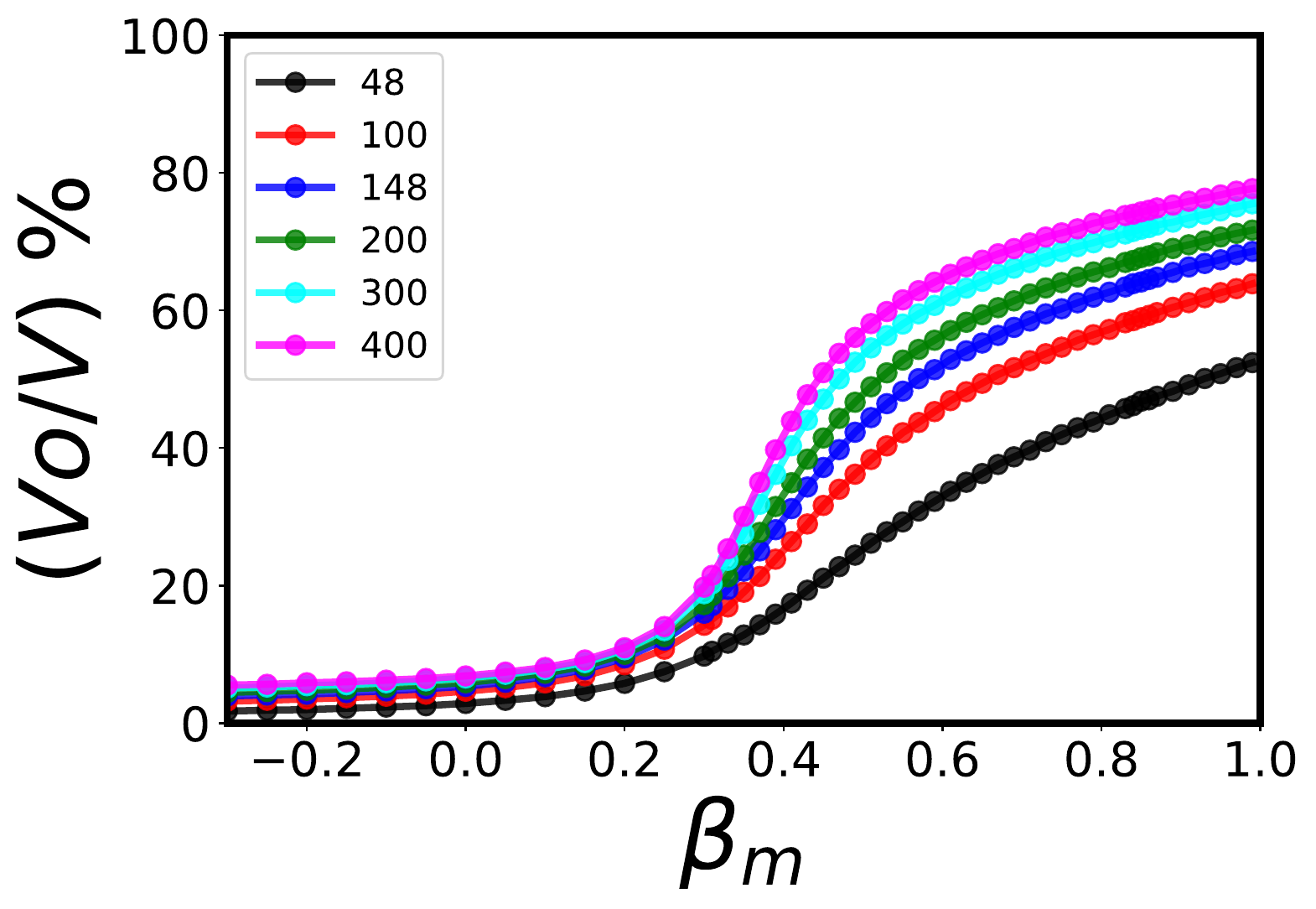}
  \caption{
Percentage of the volume fraction of the region shared by both polygons as a 
function of $\beta_m$.}
\label{fig:XXX}
\end{figure}

A natural observable for the segregated/mixed transition is based on the estimate 
of the overlap volume fraction, $V_o/V$, namely the volume of the box shared by the 
two polygons scaled by the total volume of the box containing the full 
system~\cite{orlandini2021linking}.  This is shown as a function of $\beta_m$ in 
figure~\ref{fig:XXX}.  If $\beta_m <\beta_m^*$ the overlap is 
relatively small (although not zero) and its value increases very mildly 
with $n$.  When $\beta_m >\beta_m^*$ the overlap volume fraction $V_o/V$ 
steadily increases approaching an asymptotic value that for $n=400$ approaches
the $80 \%$ of the system volume.  This indicates a very strong interpenetration 
of the two rings when the system is well inside the mixed phase ($\beta_m = 1$).

\section{Topological entanglement}
\label{sec:entangle}

\subsection{Linking probability and average linking number}
\label{sec:linkprob}

A first characterization of the topological mutual entanglement that forms in 
the system is provided by the estimate of the probability that the two polygons 
are topologically linked.  In general two disjoint simple closed curves $C_1$ and 
$C_2$ are topologically unlinked (or splittable) if there exists a homeomorphism  $H$
of $\mathbb{R}^3$ onto itself, $H: \mathbb{R}^3 \to \mathbb{R}^3$, 
such that the images $H(C_1)$ and $H(C_2)$ can be separated by a 
plane~\cite{Rolfsen:1976}.  This definition is not convenient computationally and 
we relied instead on the notion of linking based on the computation of the 
2-variable Alexander polynomial $\Delta(t,s)$ of the link diagram.  This is done 
by encoding crossings (overpasses and underpasses in a planar projection of the 
polygon pair) and calculating $\Delta(t,s)$ from the encoding.  For details 
see reference~\cite{Rolfsen:1976}.

$\Delta(t,s)$ is not a perfect invariant able to distinguish all link types,
but if we restrict ourselves to the identification of link types with minimal 
crossing number at most $7$ its resolution will be sufficient for the analysis of
the data.  The calculation of $\Delta(t,s)$ could be prohibitively costly 
if the number of crossings $n_c$ after a planar projection is very large.
This occurs in particular when the two polygons are strongly overlapping 
in the mixed phase (that is, for $\beta_m$'s sufficiently large).  The number 
of crossings was decreased by simplifying the polygons while keeping the 
topology unaltered using BFACF moves \cite{berg1981random,de1983new}
at low temperature~\cite{van1991bfacf,baiesi2014knotted}.
This reduces the system to components of close to minimal length compatible 
with the linked state.  See figure~\ref{fig:1} for some examples of simplified
configurations.  This implementation almost always reduced the number
of crossings in the projections to well below $50$, reducing the CPU time
devoted to calculating $\Delta(t,s)$.  Notice that if a component is reduced
to length $n\leq 6$, then the pair cannot be linked for geometric reasons,
so that a calculation of $\Delta(t,s)$ is not necessary.  

The calculation of $\Delta(t,s)$ proceeded by performing $100$ independent
projections onto randomly oriented planes of the simplified configurations, then
choose amongst these the projection $P$ with the least number of crossings.
$\Delta(t,s)$ is then calculated using $P$ for $t,s\in\{2,3\}$.  This gives
four values which are compared to the values computed from the explicit expression of  
$\Delta(t,s)$ for link types up to $7$ crossings (see for instance~\cite{atlas}). 
Those cases where the Alexander polynomial is not trivial but does not correspond 
to a link with $7$ or fewer mutual crossings in its minimal projection, are classified 
as \textit{complex links}. 

\begin{figure}[ht!]
  \centering
  \includegraphics[width=0.85\textwidth]{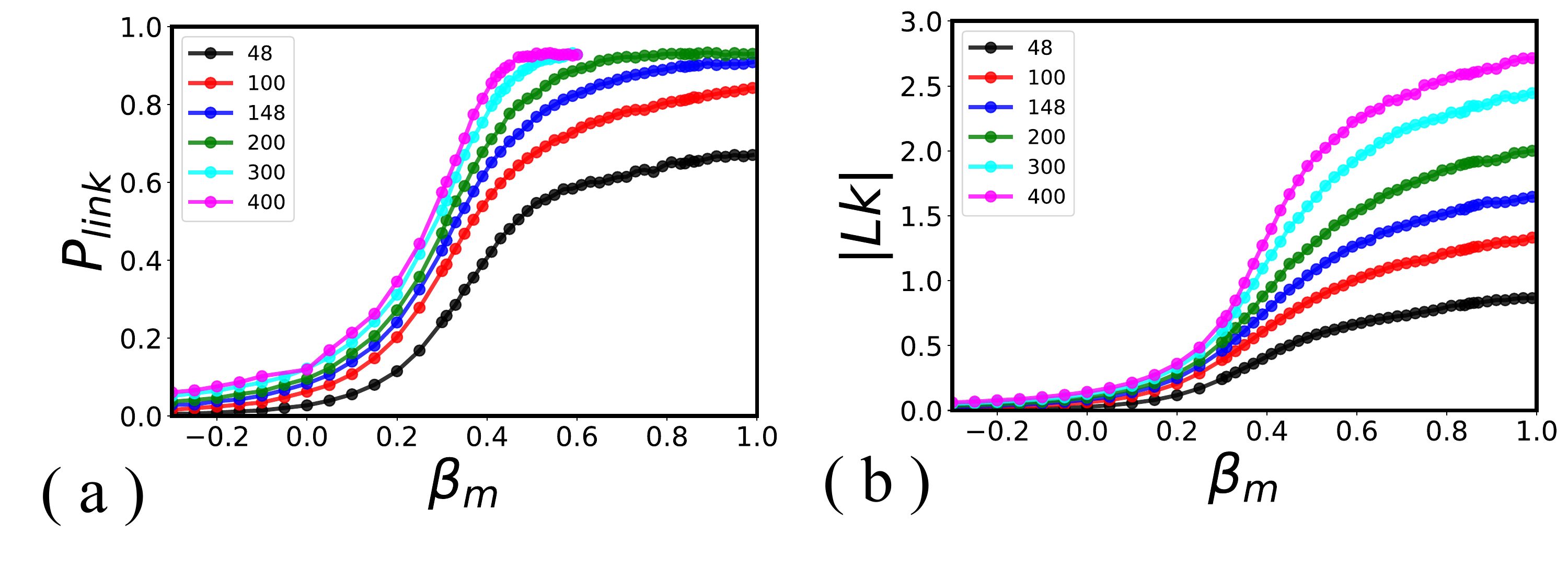}
  \caption{
(a) Probability that a pair of polygons is  linked as a function of $\beta_m$.
(b) Average of the absolute value of the linking number as a function of $\beta_m$.}
\label{fig:5}
\end{figure}

In figure~\ref{fig:5}(a) we show the probability $P_{link}$ of topologically linked pairs 
of polygons (i.e. $\Delta(t,s) \ne 0$) as a function of $\beta_m$ and for different values 
of $n$. 
There are clear qualitative trends seen in this graph and, in particular, $P_{link}$ increases
with $\beta_m$.    For $\beta_m\lessapprox 0.4$
$P_{link}$ increases rapidly with $\beta_m$ as the model transitions from the
segregated into the mixed phase.  At large values of $\beta_m$ (well inside 
the mixed phase),  $P_{link}$ appears to settle on a value close to $0.9$ at
the larger values of $n$.  
We do not give data for $n=300$ or $n=400$ when
$\beta_m\gtrapprox 0.4$ because of a possibility of false positives for the unlink.
There are link types with $\Delta(s,t)=0$ but which
are topologically linked.  These link types start to appear in the
standard knot table as having minimal crossing numbers $10$.  In our model
as $\beta_m$ increases the link types are of increasing complexity.  Computing
$\Delta(s,t)$ for some of these link types, however, gives $\Delta(s,t)=0$, and
they are classified as being the unlink.  
This causes overcounting of unlinks
in our data at large $n$ and high $\beta_m$, as well as undercounting of non-trivial links
as a consequence.  The result is that $P_{link}$ would be systematically underestimated
in figure~\ref{fig:5}(a) at large values of  $n$ and $\beta_m$.


A simpler way to measure the complexity of the linked states 
of the polygon pairs is by computing their linking number $Lk$. The linking number 
$Lk(C_1,C_2)$ of a pair of closed curves $(C_1,C_2)$ is calculated by summing 
positive and negative mutual crossings in a simple projection of 
$(C_1,C_2)$~\cite{Adams:1994}.  The linking number defines homological linking 
of $(C_1,C_2)$, namely, two curves are homologically linked if and only if 
$Lk(C1,C2)\ne 0$~\cite{Rolfsen:1976}.  In figure~\ref{fig:5}(b) we report 
the average absolute value of $Lk(C_1,C_2)$, as a  function of $\beta_m$.
These graphs of $|Lk|$ increase for all values of $\beta_m$ with $n$ and with 
increasing $\beta_m$ at fixed $n$. The increase is large in the mixed phase 
when $\beta_m>\beta_m^*$.  It shows that both increasing $n$, and increasing $\beta_m$, 
increase the complexity of the links in the mixed phase, an effect which is much less 
pronounced in the segregated phase where the probability of linking is low.

\begin{figure}[t!]
  \centering
  \includegraphics[width=0.85\textwidth]{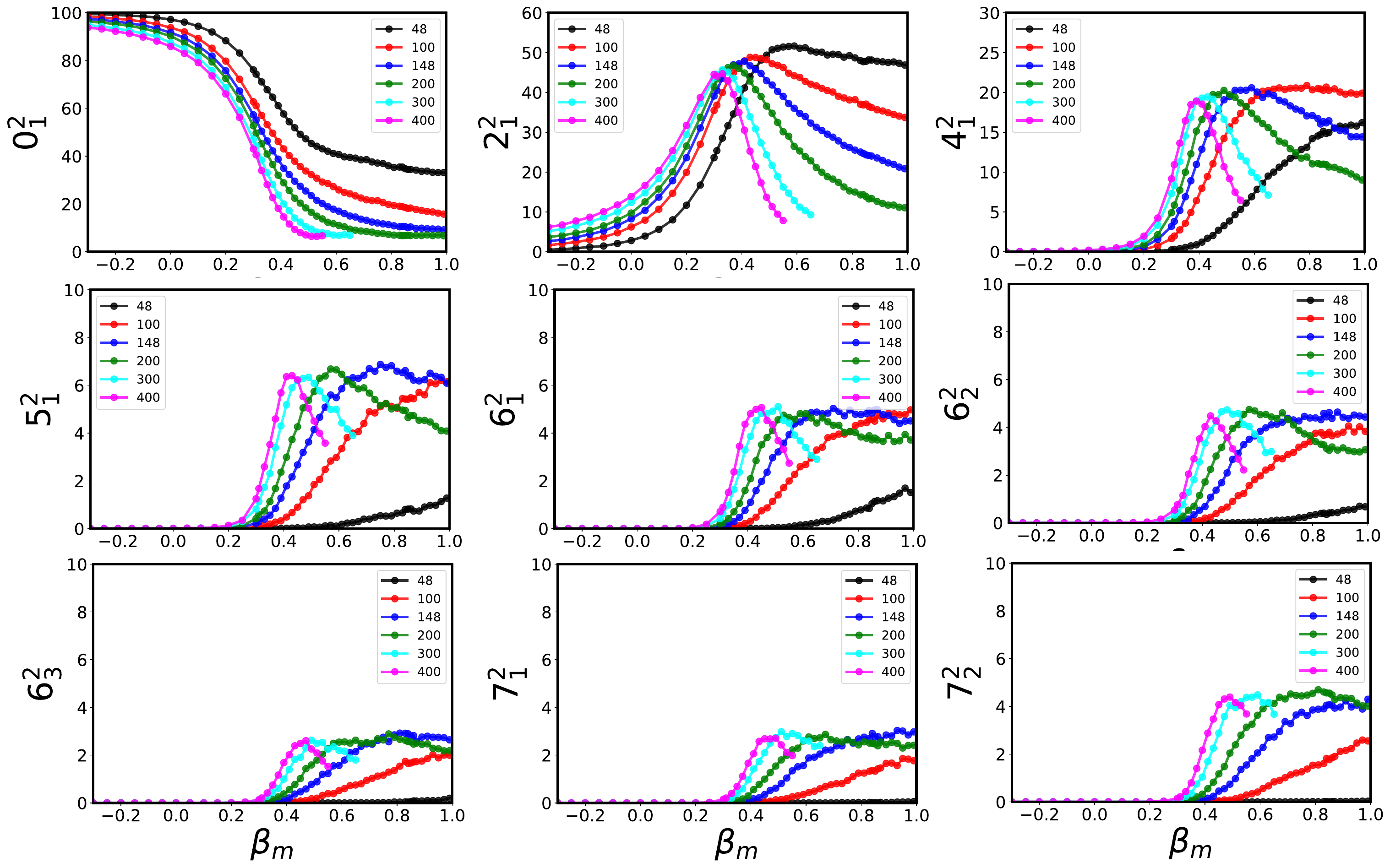}
  \caption{Percentage of the population of link types with $n_c \le 6$ as a function 
of $\beta_m$ for different values of $n$. The two link types $7_1^2$ and $7_2^2$ are the only two 
links with $n_c=7$ whose relative populations are larger than $1\%$.
}
\label{fig:6A}
\end{figure}

\subsection{Link spectrum}
\label{sec:linkspectrum}
In figure~\ref{fig:6A} we examine aspects of the link spectrum measured as
a function of $\beta_m$ by plotting the percentage of the most popular links 
(with minimal crossing number up to $7$ and with percentage at least $1\%$) as detected
in our simulations. As expected, the population of unlinks is very large for $\beta_m < 0$ 
(approaching $100$\%),  and decreases as $\beta_m$ is increased and the mixed phase 
is approached.  

For sufficiently large values of $\beta_m$ and sufficiently large values of $n$, the proportion of unlinks stabilizes 
at very low levels.  Concomitantly with this, the simplest link type (the Hopf link, $2_1^2$) 
has a non-monotonic behaviour reaching a maximum at values of $\beta_m$ that 
decrease as $n$ increases.  This non-monotonic behaviour is common to all 
link types with $n_c \le 6$ and for the $7_1^2$ and $7_2^2$ links.  The other $7$ 
crossings links also show this behaviour but their populations are too small 
(below $1\%$) for this to be significant.  

\begin{figure}[t!]
  \centering
  \includegraphics[width=0.85\textwidth]{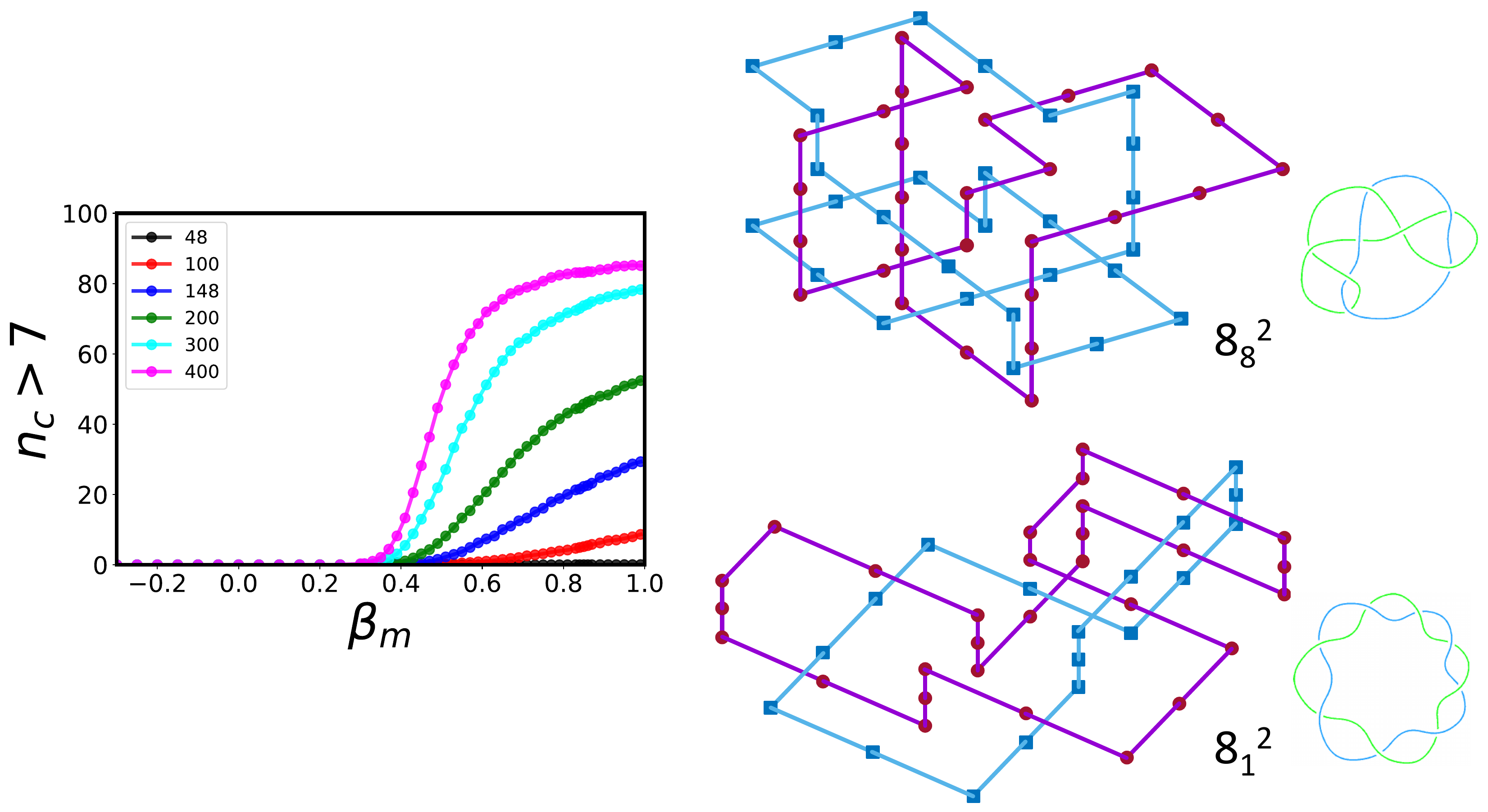}
  \caption{Percentage of the population of link types having $\Delta(t,s) \ne 0$ 
with $n_c >7$.  To give a glimpse of the complexity of the links found in the mixed 
state two examples with $n_c=8$ are reported. The top right refers to a $8^2_8$ 
link (Lk=1) and was found in a polygon pair with $n=48$ while the bottom right is 
a $8^2_1$ link (Lk=4) and was found in a polygon pair with $n=100$.  In both cases 
the configurations reported are those simplified by the smoothing algorithm based 
on BFACF moves, see text.}
\label{fig:6B}
\end{figure}

The fact that for large $n$ and well inside the mixed phase the complexity of 
the linked pairs is rapidly increasing is also suggested by the rapid increase of 
the populations of topologically linked pairs ($\Delta(t,s) \ne 0$) having $n_c >7$. 
This is reported in figure~\ref{fig:6B} together with two examples of linked pairs 
of polygons with $n_c=8$.  Again, the fact that for $n\ge 300$ the curves seem 
to approach a constant value could be due to the failure of the two-variable 
Alexander polynomial in detecting more complex links at large $\beta_m$.

\begin{figure}[t!]
  \centering
  \includegraphics[width=0.85\textwidth]{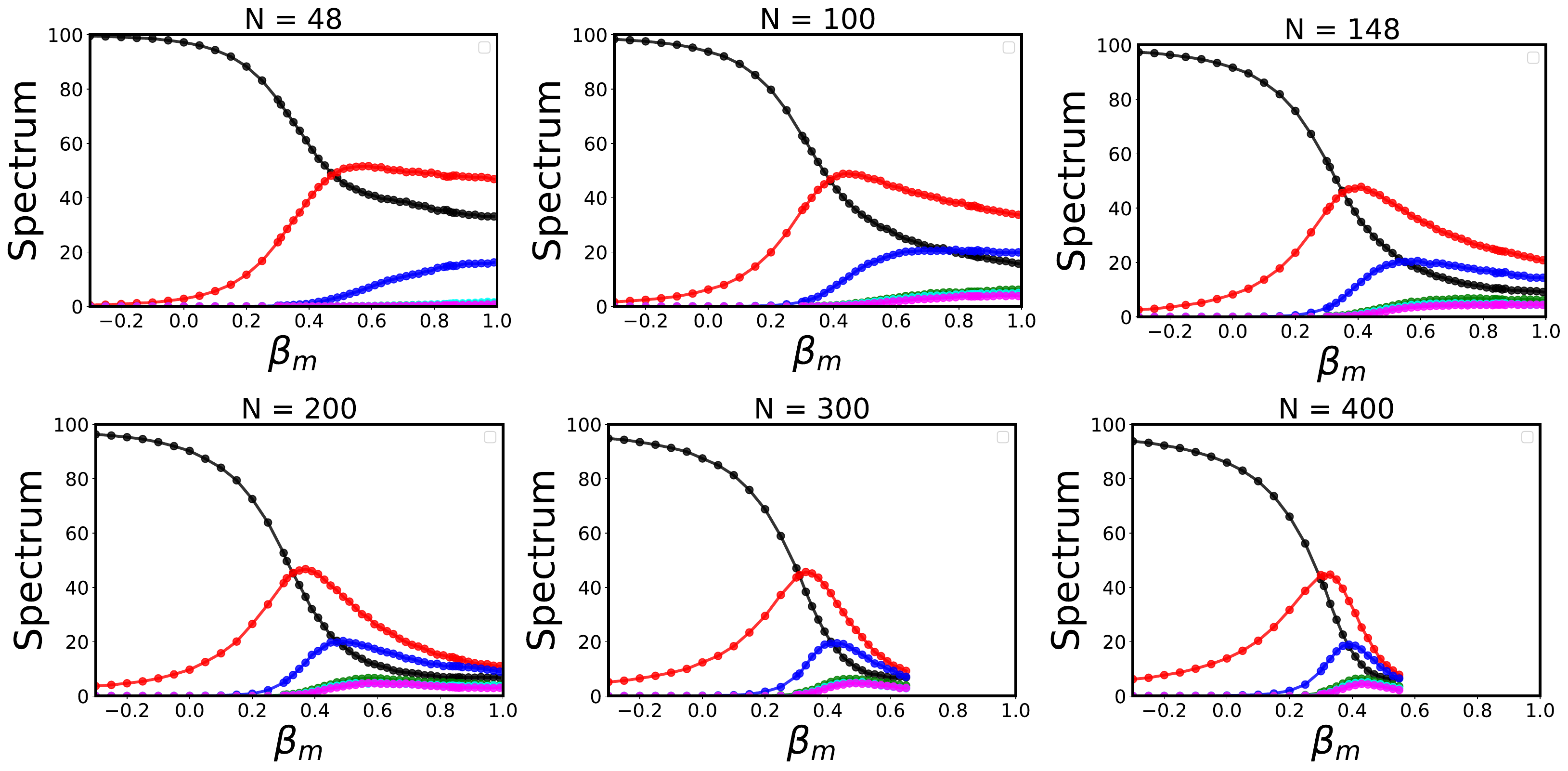}
  \caption{Percentage of the population of some link types as a function of 
  $\beta_m$ (black: $0_1^2$, red: $2_1^2$, blue: $4_1^2$, green: $5_1^2$, cyan: $6_1^2$, magenta: $6_2^2$. 
  Different panels refer to different values of $n$.}
\label{fig:7}
\end{figure}

Finally, in figure~\ref{fig:7} we report the link spectrum as a function of 
$\beta_m$ where each panel presents data for a different value of $n$. In
all cases the unlink dominates the segregated phase, but its incidence
decreases sharply when $\beta_m \approx 0.3$ while the incidence of
linked conformations increases into the mixed phase.  The Hopf link ($2_1^2$) dominates
the linked conformations in the mixed phase but it also peaks close
to $\beta_m\approx 0.3$.  More complex links also appear, albeit at
smaller proportions, as $\beta_m$ increases, and the simplest of these
similarly peak at $\beta_m\approx 0.3$.  It appears from our data that
the number of link types multiplies in the mixed phase with increasing
values of $n$, and while the incidence of specific link types decreases with
increasing $n$ and $\beta_m$, we know from figure~\ref{fig:5} that the sum 
over all these link types increases with $\beta_m$ into the mixed phase.
This may indicate that the increase in the number of 
link types compensates for the reduction in the incidence of any specific
link type, so that the proportion of linked conformations dominate state space.

\subsection{Metric and energy properties at fixed link type}

\begin{figure}[ht!]
  \centering
  \includegraphics[width=0.85\textwidth]{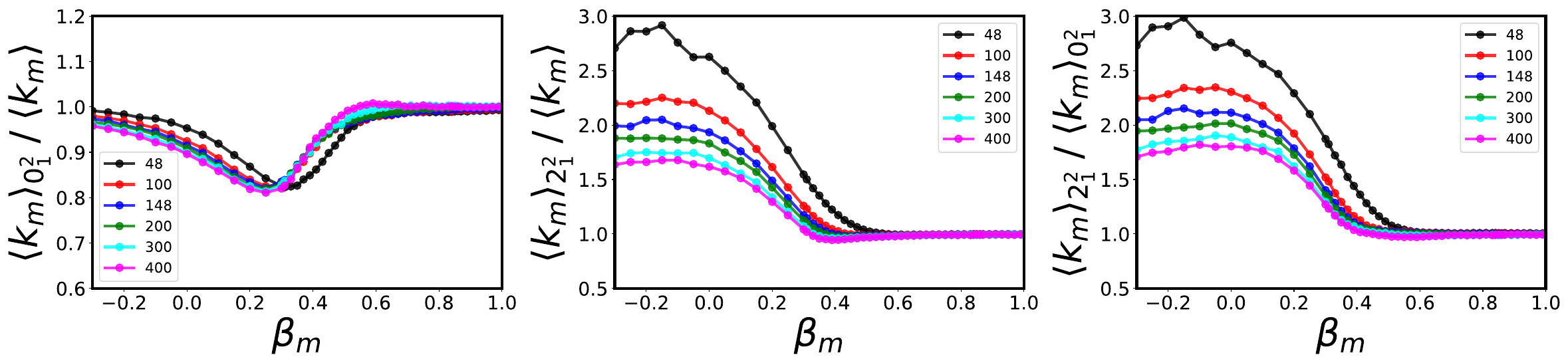}
  \caption{$\beta_m$ dependence of the ratio between  the energy of pairs of polygons 
  with a given link type and the one for  the set of all pairs of polygons or the 
  set of a different link type.  } 
\label{fig:rat_k}
\end{figure}

In figure \ref{fig:rat_k} the number of mutual contacts in pairs of unlinked polygons
is plotted as a fraction of the total number of mutual contacts.  This fraction is
(expectedly) close to one if $\beta_m$ is negative, showing that almost all conformations
are unlinked.  Increasing $\beta_m$ towards its critical value reduces this fraction, 
and this is consistent with both an increase in the total number of contacts
due the components starting to approach one another, and with an increase
in the proportion of linked conformations where the polygons are closer together
and so contain larger numbers of mutual contacts.  This observation is supported
by noting that the ratio of contacts between links of type $2^2_1$ and all
polygons in the first instance, and between $2^2_1$ and unlinked states
are high in the segregated phase, showing that linked states of type $2^2_1$
contain, on average, a higher density of mutual contacts.

Data on the mean square radius of gyration paint an interesting picture.
Deep in the segregated phase the unlink dominates state space and so the ratio
of $\langle R^2_g\rangle_{0_1^2}$ to $\langle R^2_g\rangle$ is approximately
equal to $1$.  This is also the case deep in the mixed phase -- unlinked
states have about the same size as the average conformation (since the
polygons are mixed and together collapsed into a dense conformation
minimizing the $\langle R^2_g \rangle$).  Near the critical point the 
situation is more interesting.  As the proportion of linked states increases
as $\beta_m$ approaches its critical value from below, the ratio
$\langle R^2_g\rangle_{0_1^2}/\langle R^2_g\rangle$ \textit{increases} because 
linked states are smaller than unlinked states.  This ratio should exceed $1$
(which it does).  Passing through the transition causes collapse of both
unlinked and linked states, and so the ratio should settle down to $1$
again, as it does.  This picture is  reaffirmed in the figures plotting the ratios
$\langle R^2_g\rangle_{2^2_1}/\langle R^2_g\rangle$ and
$\langle R^2_g\rangle_{2^2_1}/\langle R^2_g\rangle_{0_1^2}$ showing that
the link $2^2_1$ is \textit{larger} than the unlink and the average
of all states in the segregated phase, but are about the same size
in the mixed (collapsed) phase.

\begin{figure}[ht!]
  \centering
  \includegraphics[width=0.85\textwidth]{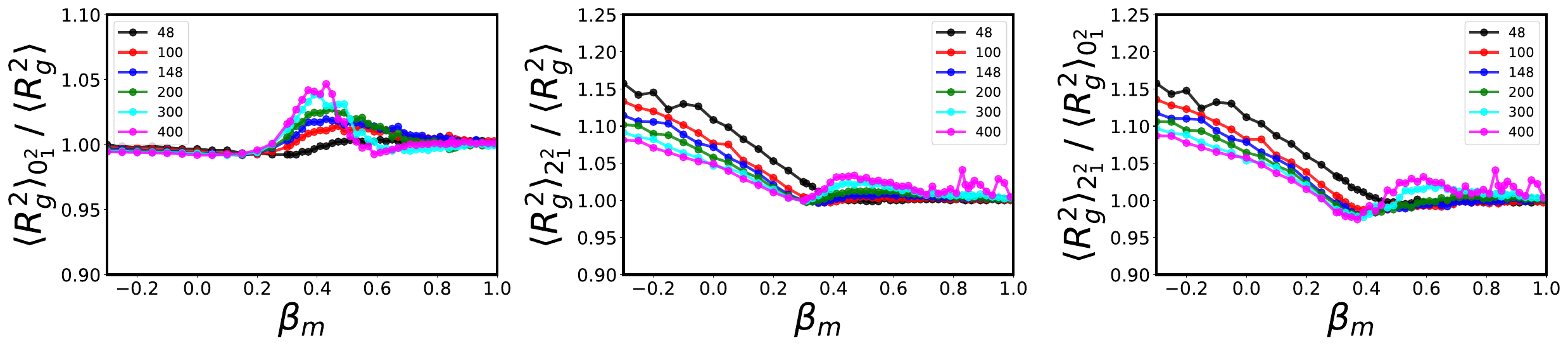}
  \caption{$\beta_m$ dependence of the ratio between  the mean-squared radius of gyration  of pairs of polygons 
of a given link type and the value for all pairs  of polygons or for pairs of a different link type. }
\label{fig:rat_Rg}
\end{figure}

\section{Discussion}
\label{sec:discussion}

To investigate the thermodynamics, metric and topological properties of a pair of 
polymer rings undergoing a segregated to mixed phase transition, 
we have considered a pair of polygons on the simple cubic lattice constrained 
to have a pair of vertices (one from each polygon) unit distance apart.  
The polygons are self- and mutually avoiding and, in addition, there is a 
short range potential between pairs of vertices in the two polygons.  When 
this potential is repulsive or weakly attractive  the two polygons are largely 
separated in space but when the potential is sufficiently attractive the 
polygons interpenetrate and form a more compact object in a mixed phase.

In section \ref{sec:rigorous} we prove that the limiting free energy exists when 
the potential is repulsive, and we establish bounds when it is attractive that establish 
the existence of a phase transition from a segregated phase to one where there are many 
inter-polygon contacts.  We use a Monte Carlo approach to investigate configurational 
properties such as the expected number of inter-polygon contacts, and the radius of 
gyration of a polygon as a function of the strength of the potential.  The mean number 
of contacts increases as the potential becomes more attractive and increases rapidly 
in the region of the transition.  The radius of gyration scales differently (with size) in 
the segregated and compact phases and there are changes in the asphericity and prolateness.

In section \ref{sec:entangle} we looked at the extent of linking of the two polygons 
as a function of the strength of the potential, both by computing the 2-variable 
Alexander polynomial (as a detector of topological linking) and the linking number 
(as a detector of homological linking).  As the potential becomes more attractive 
the linking probability and the link complexity both increase, with a relatively sharp 
increase around the transition region.  It is clear that, in the compact phase where 
there is considerable interpenetration of the polygons, the linking probability is high.

A related experimental situation is as follows.  Consider a uniform 4-star polymer with two $A$-arms and two $B$-arms
with the ends of the arms functionalized so that the system can be cyclized to form 
an $A$ ring and a $B$ ring, forming a figure eight.  The $A$ and $B$ arms carry opposite charges so that they 
are attracted to one another and the strength of the attraction can be modified by changing the pH or the ionic strength.
Prepare the system (\emph{ie} the 4-star) at some fixed pH and ionic strength and, after equilibration, carry
out a cyclization reaction.  At high ionic strength or where charges are suppressed by varying the 
pH, the $A$ and $B$ arms will repel or weakly attract and there should be little interpenetration so
that, after cyclization, there should be little linking.  Conversely, with large charge densities and low ionic strength
there should be considerable interpenetration and linking.

\section*{Acknowledgement}
EJJvR acknowledges financial support from NSERC (Canada) 
in the form of a Discovery Grant RGPIN-2019-06303.   Data generated for this study
are available on reasonable request.

\bibliography{biblio.bib}

\providecommand{\newblock}{}
\begin{thebibliography}{10}
\expandafter\ifx\csname url\endcsname\relax
  \def\url#1{{\tt #1}}\fi
\expandafter\ifx\csname urlprefix\endcsname\relax\def\urlprefix{URL }\fi
\providecommand{\eprint}[2][]{\url{#2}}

\bibitem{DeGennes:1979}
de~Gennes P~G 1979 {\em Scaling concepts in Polymer Physics\/} (Cornell
  University Press, Ithaca, New York)

\bibitem{van1994lattice}
Janse~van Rensburg E~J, Orlandini E, Sumners D~W, Tesi M~C and Whittington S~G
  1994 {\em Phys. Rev. E\/} {\bf 50} R4279

\bibitem{janse1996entanglement}
Janse~van Rensburg E~J, Orlandini E, Sumners D~W, Tesi M~C and Whittington S~G
  1996 {\em J. Stat. Phys.\/} {\bf 85} 103--130

\bibitem{hammersley1961number}
Hammersley J~M 1961 {\em Math. Proc. Camb. Phil. Soc.\/} {\bf 57} 516--523

\bibitem{Pippenger:1989:DAM}
Pippenger N 1989 {\em Discr. Appl. Math.\/} {\bf 25} 273--278

\bibitem{Sumners&Whittington:1988:J-Phys-A}
Sumners D~W and Whittington S~G 1988 {\em J. Phys. A: Math. Gen.\/} {\bf 21}
  1689--1694

\bibitem{soteros1992entanglement}
Soteros C~E, Sumners D~W and Whittington S~G 1992 {\em Math. Proc. Camb. Phil.
  Soc.\/} {\bf 111} 75--91

\bibitem{van2002probability}
Janse~van Rensburg E~J 2002 {\em Contemporary Mathematics\/} {\bf 304} 125--136

\bibitem{baiesi2010entropic}
Baiesi M, Orlandini E and Stella A~L 2010 {\em J. Stat. Mech.: Theo. Exp.\/}
  {\bf 2010} P06012

\bibitem{madras1990monte}
Madras N, Orlitsky A and Shepp L~A 1990 {\em J.Stat. Phys.\/} {\bf 58} 159--183

\bibitem{verdier1962monte}
Verdier P~H and Stockmayer W~H 1962 {\em J. Chem. Phys.\/} {\bf 36} 227--235

\bibitem{tesi1996monte}
Tesi M~C, Janse~van Rensburg E~J, Orlandini E and Whittington S~G 1996 {\em J.
  Stat. Phys.\/} {\bf 82} 155--181

\bibitem{tesi1996interacting}
Tesi M~C, Janse~van Rensburg E~J, Orlandini E and Whittington S~G 1996 {\em J.
  Phys. A: Math. Gen.\/} {\bf 29} 2451

\bibitem{Geyer:1991:CSS}
Geyer C~J 1991 {\em Comp. Sci. and Stat.: Proc. 23rd Symp. on the Interface\/}
  156--163

\bibitem{Clisby:2010:Phys-Rev-Lett}
Clisby N 2010 {\em Phys. Rev. Lett.\/} {\bf 104} 055702

\bibitem{orlandini2021linking}
Orlandini E, Tesi M~C and Whittington S~G 2021 {\em J. Phys. A: Math. Theo.\/}
  {\bf 54} 505002

\bibitem{Rolfsen:1976}
Rolfsen D 1976 {\em Knots and Links\/} (Mathematics Lecture Series 7, Publish
  or Perish, Inc., Houston, Texas)

\bibitem{berg1981random}
Berg B and Foerster D 1981 {\em Phys. Lett. B\/} {\bf 106} 323--326

\bibitem{de1983new}
Aragao De~Carvalho C and Caracciolo S 1983 {\em J. de Physique\/} {\bf 44}
  323--331

\bibitem{van1991bfacf}
Janse~van Rensburg E~J and Whittington S~G 1991 {\em J. Phys. A: Math. Gen.\/}
  {\bf 24} 5553

\bibitem{baiesi2014knotted}
Baiesi M, Orlandini E and Stella A~L 2014 {\em Macromol.\/} {\bf 47} 8466--8476

\bibitem{atlas}
The knot atlas \url{http://katlas.org/wiki/Main_Page}

\bibitem{Adams:1994}
Adams C~C 1994 {\em The Knot Book\/} (Freeman)

\end{thebibliography}

\end{document}